\documentclass[twocolumn]{aastex63}
\usepackage{silence} 
\usepackage{times}
\usepackage{enumitem}
\usepackage{graphicx}
\usepackage{amsmath}
\usepackage{bm}
\usepackage{graphics}
\usepackage{url}
\usepackage{epsfig}
\usepackage{wrapfig}
\usepackage{ulem}
\usepackage{verbatim}
\usepackage{float}
\usepackage{lastpage}
\usepackage{longtable}
\usepackage{xspace}
\usepackage{natbib}

\makeatletter
\define@key{Gin}{resolution}{\pdfimageresolution=#1\relax}
\makeatother

\providecommand{\apjl}{ApJ}

\providecommand{\apjs}{ApJS}
\providecommand{\aap}{A\&A}

\def\simlt{\mathrel{\hbox{\rlap{\hbox{\lower4pt\hbox{$\sim$}}}\hbox{$<$}}}}
\def\simgt{\mathrel{\hbox{\rlap{\hbox{\lower4pt\hbox{$\sim$}}}\hbox{$>$}}}}

\def\I{\,\textsc{i}}

\def\V{\,\textsc{v}}

\def\arcsec{$^{\,\prime\prime}$}

\def\arcdeg{\degr}

\def\Ncandidate{290} 
\def\Ngaia{13}       
\def\Nmpc{2}         
\def\Nclass{31}      
\def\Nvariable{3}    
\def\Nredshift{183}  
\def\Nbeforephot{58}
\def\Nphot{30}       
\def\Nremain{28}     

\def\teglon{{\tt Teglon}}

\def\ptDi{P_{{\rm 2D}_{i}}}

\def\ppptDi{P^{''}_{{\rm 2D}_{i}}}

\def\finalninetyarea{$9,881$~deg$^{2}$}
\def\initialninetyarea{$10,183$~deg$^{2}$}
\def\finalfiftyarea{$2,400$~deg$^{2}$}
\def\fourDninetyarea{$6,674$~deg$^{2}$}

\def\panstarrsarea{$1,085$~deg$^{2}$\xspace}
\def\atlasarea{$7,110$~deg$^{2}$\xspace}

\def\lcogtpub{Keinan et al. {\it in prep.}}
\def\coulterpub{Coulter 2024, {\it in prep.}}

\shorttitle{The Gravity Collective: GW190425}
\shortauthors{Coulter et~al.}

\begin{document}

\title{The Gravity Collective: A Comprehensive Analysis of the Electromagnetic Search for the Binary Neutron Star Merger GW190425}

\newcommand{\NU}{\affiliation{Center for Interdisciplinary Exploration and Research in Astrophysics (CIERA), Northwestern University, Evanston, IL 60201, USA}}
\newcommand{\fermi}{\affiliation{Fermi National Accelerator Laboratory, P. O. Box 500, Batavia, IL 60510, USA}}
\newcommand{\austinstate}{\affiliation{Austin Peay State University, Dept. Physics, Engineering and Astronomy, P.O. Box 4608, Clarksville, TN 37044, USA}}
\newcommand{\UCSC}{\affiliation{Department of Astronomy and Astrophysics, University of California, Santa Cruz, CA 95064, USA}}
\newcommand{\carnegie}{\affiliation{The Observatories of the Carnegie Institution for Science, 813 Santa Barbara St., Pasadena, CA 91101, USA}}
\newcommand{\kipac}{\affiliation{Kavli Institute for Particle Astrophysics \& Cosmology, P. O. Box 2450, Stanford University, Stanford, CA 94305, USA}}
\newcommand{\LBNL}{\affiliation{Lawrence Berkeley National Laboratory, 1 Cyclotron Road, MS 50B-4206, Berkeley, CA 94720-3411, USA}}
\newcommand{\benedictine}{\affiliation{Benedictine University, Department of Physics, 5700 College Road, Lisle, IL, 60532, USA}}
\newcommand{\ets}{\affiliation{Department of Physics and Astronomy, East Tennessee State University, Johnson City, TN 37614, USA}}
\newcommand{\CBPF}{\affiliation{Centro Brasileiro de Pesquisas F\'isicas, Rua Dr. Xavier Sigaud 150, CEP 22290-180, Rio de Janeiro, RJ, Brazil}}
\newcommand{\CEFET}{\affiliation{Centro Federal de Educa\c{c}\~ao Tecnol\'ogica Celso Suckow da Fonseca, Rodovia M\'ario Covas, lote J2, quadra J, CEP 23810-000,  Itagua\'i, RJ, Brazil}}
\newcommand{\ICAS}{\affiliation{International Center for Advanced Studies \& Instituto de Ciencias Físicas,  ECyT-UNSAM \& CONICET, 1650, Buenos Aires, Argentina}}
\newcommand{\NATU}{\affiliation{NAT-Universidade Cruzeiro do Sul / Universidade Cidade de S{\~a}o Paulo, Rua Galv{\~a}o Bueno, 868, 01506-000, S{\~a}o Paulo, SP, Brazil}}
\newcommand{\IITH}{\affiliation{Indian Institute of Technology, Hyderabad, Kandi Telangana 502285 India}}
\newcommand{\UCBerkeley}{\affiliation{Department of Astronomy, University of California, Berkeley, CA 94720-3411, USA}}
\newcommand{\Miller}{\affiliation{Miller Institute for Basic Research in Science, University of California, Berkeley, CA 94720, USA}}
\newcommand{\IAPUCC}{\affiliation{Instituto de Astrof\'isica, Pontificia Universidad Cat\'olica de Chile, Casilla 306, Santiago 22, Chile}}
\newcommand{\millennium}{\affiliation{Millennium Institute of Astrophysics (MAS), Nuncio Monse$\tilde{n}$or S\'otero Sanz 100, Providencia, Santiago, Chile}}
\newcommand{\carnegieLCO}{\affiliation{Carnegie Observatories, Las Campanas Observatory, Casilla 601, La Serena, Chile}}
\newcommand{\CCNE}{\affiliation{Departamento de F\'isica, Centro de Ci\^encias Naturais e Exatas, Universidade Federal de Santa Maria, 97105-900, Santa Maria, RS, Brazil}}
\newcommand{\UDP}{\affiliation{N\'ucleo de Astronom\'ia, Universidad Diego Portales, Av. Ejército 441, Santiago, Chile}}
\newcommand{\Michigan}{\affiliation{Department of Physics, University of Michigan, Ann Arbor, MI 48109, USA}}
\newcommand{\UWMadison}{\affiliation{Physics Department, University of Wisconsin-Madison, Madison, WI 53706, USA}}
\newcommand{\Steward}{\affiliation{Steward Observatory, University of Arizona, 933 North Cherry Avenue, Tucson, AZ 85721-0065, USA}}
\newcommand{\TelAviv}{\affiliation{The School of Physics and Astronomy, Tel Aviv University, Tel Aviv 69978, Israel}}
\newcommand{\CIFAR}{\affiliation{CIFAR Azrieli Global Scholars program, CIFAR, Toronto, Canada}}
\newcommand{\MtStromlo}{\affiliation{Mt Stromlo Observatory, The Research School of Astronomy and Astrophysics, Australian National University, ACT 2601, Australia}}
\newcommand{\NCPAS}{\affiliation{National Centre for the Public Awareness of Science, Australian National University, Canberra, ACT 2611, Australia}}
\newcommand{\ARC}{\affiliation{The ARC Centre of Excellence for All-Sky Astrophysics in 3 Dimensions (ASTRO 3D), Australia}}
\newcommand{\ColumbiaAL}{\affiliation{Columbia Astrophysics Laboratory, Columbia University, New York, NY 10027, USA}}
\newcommand{\flatiron}{\affiliation{Center for Computational Astrophysics, Flatiron Institute, 162 W. 5th Avenue, New York, NY 10011, USA}}
\newcommand{\IU}{\affiliation{Department of Astronomy, Indiana University, 727 E. Third St., Bloomington, IN 47405}}
\newcommand{\Racah}{\affiliation{Racah Institute for Physics, The Hebrew University, Jerusalem Israel 91904}}
\newcommand{\MPIA}{\affiliation{Max-Planck-Institut fur Astrophysik, Karl-Schwarzschild-Str 1, D-85748 Garching bei M\"unchen, Germany}}
\newcommand{\UCSB}{\affiliation{Department of Physics, University of California, Santa Barbara, CA 93106-9530, USA}}
\newcommand{\LCO}{\affiliation{Las Cumbres Observatory, 6740 Cortona Dr, Suite 102, Goleta, CA 93117-5575, USA}}
\newcommand{\efermi}{\affiliation{Enrico Fermi Institute, Department of Physics, Department of Astronomy and Astrophysics}}
\newcommand{\kavlicosmo}{\affiliation{Enrico Fermi Institute, Department of Physics, Department of Astronomy and Astrophysics,\\and Kavli Institute for Cosmological Physics, University of Chicago, Chicago, IL 60637, USA}}
\newcommand{\jordell}{\affiliation{Jodrell Bank Centre for Astrophysics, University of Manchester, Oxford Road, Manchester, UK}}
\newcommand{\UCD}{\affiliation{Department of Physics and Astronomy, University of California, Davis, 1 Shields Avenue, Davis, CA 95616-5270, USA}}
\newcommand{\IFT}{\affiliation{Instituto de F\'isica Te\'orica UAM/CSIC, Universidad Aut\'onoma de Madrid, 28049 Madrid, Spain}}
\newcommand{\CAS}{\affiliation{Centre for Astrophysics and Supercomputing, Swinburne University of Technology, PO Box 218, H29, Hawthorn, VIC, 3122, Australia}}
\newcommand{\OzGrav}{\affiliation{Australian Research Council Centre of Excellence for Gravitational Wave Discovery, Swinburne University of Technology, Hawthorn, VIC, 3122, Australia}}
\newcommand{\UAB}{\affiliation{Departamento de Ciencias Fisicas, Universidad Andres Bello, Avda. Republica 252, Santiago, Chile}}
\newcommand{\UQB}{\affiliation{School of Mathematics and Physics, University of Queensland, Brisbane, QLD, 4072, Australia}}
\newcommand{\weizmann}{\affiliation{Department of Particle Physics and Astrophysics, Weizmann Institute of Science, Rehovot, 7610001, Israel}}
\newcommand{\UChicago}{\affiliation{Department of Astronomy and Astrophysics, University of Chicago, Chicago, IL 60637, USA}}
\newcommand{\atacama}{\affiliation{Instituto de Astronom\'{\i}a y Ciencias Planetarias, Universidad de Atacama, Copayapu 485, Copiap\'o, Chile}}
\newcommand{\Thacher}{\affiliation{Thacher Observatory, Thacher School, 5025 Thacher Rd. Ojai, CA 93023, USA}}
\newcommand{\Hawaii}{\affiliation{Institute for Astronomy, University of Hawaii, 2680 Woodlawn Drive, Honolulu, HI 96822, USA}}
\newcommand{\camino}{\affiliation{Departamento de Astronom\'ia, Universidad de Chile, Camino El Observatorio 1515, Las Condes, Santiago, Chile}}
\newcommand{\JHU}{\affiliation{Department of Physics and Astronomy, Johns Hopkins University, 3400 North Charles Street, Baltimore, MD 21218, USA}}
\newcommand{\STScI}{\affiliation{Space Telescope Science Institute, 3700 San Martin Drive, Baltimore, MD 21218, USA}}
\newcommand{\mcgill}{\affiliation{Department of Physics, McGill University, 3600 University Street, Montr\'eal, QC H3A 2T8, Canada}}
\newcommand{\mcgillinst}{\affiliation{McGill Space Institute, McGill University, 3550 University Street, Montr\'eal, QC H3A 2A7, Canada}}
\newcommand{\cdita}{\affiliation{Center for Data Intensive and Time Domain Astronomy, Department of Physics and Astronomy, Michigan State University, East Lansing, MI 48824, USA}}
\newcommand{\cfa}{\affiliation{Center for Astrophysics, Harvard \& Smithsonian, 60 Garden St, Cambridge, MA 02138, USA}}
\newcommand{\Einstein}{\affiliation{NASA Einstein Fellow}}

\newcommand{\Wesleyan}{\affiliation{Department of Mathematics and Computer Science, Wesleyan University, 45 Wyllys Ave, Middletown, CT 06457, USA}}

\newcommand{\CITEVA}{\affiliation{Astronomy Center (CITEVA), University of Antofagasta, Avenida U. de Antofagasta, 02800 Antofagasta, Chile}}

\newcommand{\trinity}{\affiliation{School of Physics, Trinity College Dublin, The University of Dublin, Dublin 2, Ireland}}

\newcommand{\llnl}{\affiliation{Space Science Institute, Lawrence Livermore National Laboratory, 7000 East Avenue, Livermore, CA 94550, USA}}

\newcommand{\tasmania}{\affiliation{University of Tasmania, Physics, UTAS Physics Building - Private Bag 37, Hobart, Tasmania, 7001, Australia}}

\newcommand{\NARIT}{\affiliation{National Astronomical Research Institute of Thailand, 260 Moo 4, Donkaew, Maerim, Chiang Mai, 50180, Thailand}}

\newcommand{\GSFC}{\affiliation{Astrophysics Science Division, NASA Goddard Space Flight Center, Greenbelt, MD 20771, USA}}
\newcommand{\JSSI}{\affiliation{Joint Space-Science Institute, University of Maryland, College Park, MD 20742, USA }}

\newcommand{\dunlap}{\affiliation{David A. Dunlap Department of Astronomy and Astrophysics, University of Toronto, 50 St. George Street, Toronto, Ontario, M5S 3H4, Canada}}

\correspondingauthor{D.~A.~Coulter}
\email{dcoulter@stsci.edu}

\author[0000-0003-4263-2228]{D.~A.~Coulter}
\STScI
\UCSC

\author[0000-0002-5740-7747]{C.~D.~Kilpatrick}
\NU

\author[0000-0002-6230-0151]{D.~O.~Jones}
\Hawaii

\author[0000-0002-2445-5275]{R.~J.~Foley}
\UCSC

\author[0000-0001-9051-1338]{J.~Anais~Vilchez}
\CITEVA

\author[0000-0001-7090-4898]{I. Arcavi}
\TelAviv

\author[0000-0001-8756-1262]{K.~E.~Clever}
\UCSC

\author[0000-0001-9494-179X]{G.~Dimitriadis}
\trinity

\author[0000-0003-3460-0103]{A.~V.~Filippenko}
\UCBerkeley

\author{N. Mu\~noz-Elgueta}
\MPIA

\author[0000-0001-6806-0673]{A.~L.~Piro}
\carnegie

\author[0000-0003-4131-1676]{P.~J.~Qui\~ nonez}
\UCSC

\author{G.~S.~Rahman}
\Thacher
\Wesleyan

\author[0000-0002-7559-315X]{C.~Rojas-Bravo}
\UCSC

\author[0000-0003-2445-3891]{M.~R.~Siebert}
\STScI

\author[0000-0001-9316-5389]{H.~E.~Stacey}
\Thacher

\author[0000-0002-9486-818X]{J.~J.~Swift}
\Thacher

\author[0000-0002-2636-6508]{W.~Zheng}
\UCBerkeley

\author[0000-0002-7777-216X]{J.\ S.\ Bloom}
\UCBerkeley

\author[0000-0003-0416-9818]{M.~J.~Bustamante-Rosell}
\UCSC

\author[0000-0002-5680-4660]{K.~W.~Davis}
\UCSC

\author[0009-0004-7605-8484]{J.~Kutcka}
\UCSC

\author[0000-0002-9946-4635]{P.~Macias}
\UCSC

\author[0000-0002-1052-6749]{P.~McGill}
\UCSC
\llnl

\author[0000-0003-2558-3102]{E.~Ramirez-Ruiz}
\UCSC

\author{K.~Siellez}
\tasmania

\author[0000-0002-1481-4676]{S.~Tinyanont}
\NARIT

\author[0000-0003-1673-970X]{S.~B.~Cenko}
\GSFC
\JSSI

\author[0000-0001-7081-0082]{M.~R.~Drout}
\dunlap

\author[0000-0002-8543-761X]{R.~Hausen}
\JHU

\author[0000-0003-4253-656X]{D.~Andrew Howell}
\LCO\UCSB

\author[0000-0002-3934-2644]{W.~V.~Jacobson-Gal\'{a}n}
\UCBerkeley

\author[0000-0002-5981-1022]{D.~Kasen}
\UCBerkeley\LBNL

\author[0000-0001-5807-7893]{C.~McCully}
\LCO

\author[0000-0002-4410-5387]{A.~Rest}
\STScI\JHU

\author[0000-0002-5748-4558]{K.~Taggart}
\UCSC

\author[0000-0001-8818-0795]{S.~Valenti}
\UCD

\begin{abstract}
We present an ultraviolet-to-infrared search for the electromagnetic (EM) counterpart to GW190425, the second-ever binary neutron star (BNS) merger discovered by the LIGO-Virgo-KAGRA Collaboration (LVK).  GW190425 was more distant and had a larger localization area than GW170817, therefore we use a new tool \teglon\ to redistribute the GW190425 localization probability in the context of galaxy catalogs within the final localization volume. We derive a 90th percentile area of 6,688~deg$^{2}$, a $\sim$1.5$\times$ improvement relative to the LIGO/Virgo map, and show how \teglon~provides an order of magnitude boost to the search efficiency of small ($\leq$1~deg$^{2}$) field-of-view instruments. We combine our data with all publicly reported imaging data, covering 9,078.59 deg$^2$ of unique area and 48.13\% of the LIGO/Virgo-assigned localization probability, to calculate the most comprehensive kilonova, short gamma-ray burst (sGRB) afterglow, and model-independent constraints on the EM emission from a hypothetical counterpart to GW190425 to date under the assumption that no counterpart was found in these data. If the counterpart were similar to AT~2017gfo, there was a 28.4\% chance that it would have been detected in the combined dataset. We are relatively insensitive to an on-axis sGRB, and rule out a generic transient with a similar peak luminosity and decline rate as AT~2017gfo to $30$\% confidence.  Finally, across our new imaging and all publicly-reported data, we find 28 candidate optical counterparts that we cannot rule out as being associated with GW190425, finding that 4 such counterparts discovered within the localization volume and within 5~days of merger exhibit luminosities consistent with a kilonova. 

\end{abstract}


\section{Introduction}\label{sec:intro}

The mergers of neutron stars (NSs) and black holes (BHs) produce sufficiently strong gravitational waves (GWs) that they can be detected by modern interferometric instruments such as the Laser Interferometer Gravitational-wave Observatory (LIGO) and Virgo \citep{Abbott16:gw,Abbott16:GW151226,Abbott17,2017ApJ...851L..35A,Abbott17:0814}.  The majority of detected GW events involve binary black holes \citep[BBHs;][]{GWTC3}, systems that are naively expected to produce no electromagnetic (EM) emission.  However, the LIGO-Virgo-KAGRA (LVK) collaboration has detected nine mergers of compact binaries where at least one component has a mass consistent with being an NS \citep{O3a, GWTC3,LIGO:GW230529}.  In these cases, there is the potential for an electromagnetically luminous counterpart, such as as a short gamma-ray burst (sGRB) or a radioactively powered kilonova \citep[KN;][]{Li98,Shibata06,Metzger10,Roberts2011,Kasen17}.

A single GW event, GW170817 \citep{Abbott17:detection}, has been observed electromagnetically as GRB\,170817A \citep{Goldstein17,Savchenko17} and SSS17a/AT~2017gfo \citep{Coulter17}.  This event was the result of the merger of two roughly equal-mass NSs with component masses of 1.46$\substack{+0.12\\-0.10}$ and 1.27$\pm$0.09~M$_{\sun}$ and a total mass of 2.73$\substack{+0.04\\-0.01}$~M$_{\sun}$ \citep{Abbott17:detection}.  The ultraviolet (UV), optical, and infrared (IR; collectively denoted as UVOIR) data are consistent with a radioactively powered KN with 0.06~M$_{\sun}$ of ejecta that is rich in $r$-process material \citep{Arcavi17,Cowperthwaite17,Drout17,Kasliwal17,Kilpatrick17,Smartt17,Soares-Santos17,Valenti17}.  The GRB and its afterglow, observed as a nonthermal component for several years \citep{Haggard17,Margutti17,Murguia-Berthier17,Alexander18,Lyman18,Margutti18,Nynka18,Pooley18,Ruan18,Troja18,Fong19,Hajela19,Piro19,Troja19,Troja20,Murguia-Berthier2021,Makhathini21,Hajela22,Kilpatrick22}, are consistent with a structured jet having an opening angle of $\sim 5^{\circ}$ pointed $\sim 20^{\circ}$ from our line of sight.

Localizing new EM counterparts to GW events has been a major focus of GW astronomy since 2017.  While GW events up through LVK observing run 4a \citep[O4a;][]{O3a, GWTC3,LIGO:GW230529} have been localized to a precision of, at best, tens of square degrees, arcsecond localizations of their EM counterparts are necessary to enable analysis of their host environments, the mechanisms that power their GRB and KN emission, and new studies in cosmology and NS physics that require both GW and EM emission.  For example, simultaneous detection of NS mergers in GW and EM emission have led to new constraints on the nuclear equation of state \citep{Capano20}, studies of the nature of gravity \citep{Baker17}, analysis of NS merger populations in the local Universe to compare with GRBs discovered at redshift $z>0.1$ \citep{Fong17,Fong22,Nugent22}, a novel method to measure the Hubble constant \citep{Abbott17:cosmo}, and analysis of the sites and mechanisms for $r$-process production \citep{Cowperthwaite17,Drout17,Kasliwal17,Kilpatrick17,Smartt17}.

Of the eight compact-binary mergers that the LVK has detected through its third observing run (O3), five have one component that is consistent with that of a BH (i.e., a neutron-star--black-hole merger, NSBH) and a mass ratio where the secondary component (if an NS) is expected to be disrupted inside the innermost stable circular orbit, precluding any EM emission.

Besides GW170817, the only other BNS merger yet detected is GW190425, which consisted of a 2.02$\substack{+0.58\\-0.34}$ and 1.35$\pm$0.26~M$_{\odot}$ NS with a total mass of 3.4$\substack{+0.3\\-0.1}$~M$_{\odot}$ \citep{Abbott20:gw190425}. Unfortunately, GW190425 was a ``single-detector'' event, only observed by the LIGO Livingston detector.  Consequently, its initial (final) localization was constrained to \hbox{\initialninetyarea}~(\hbox{\finalninetyarea}) at 90\% confidence, covering roughly one quarter of the sky.  Additionally, its high total mass implies a KN that is fainter and redder than AT~2017gfo \citep{Foley20}.  Because a large fraction of the localization region was close to the Sun, no observatory could practically observe the entire localization region.  Moreover, the size of the localization region and its extent over both hemispheres meant multiple telescopes were necessary to cover the maximum area possible.

Starting 15.5~hr after the trigger, we observed portions of the GW190425 localization region using five small-aperture telescopes as part of the One-Meter, Two-Hemispheres (1M2H) team.  At the same time, several other teams, including GRANDMA, GROWTH, GOTO-4, SAGUARO, and others, began their own observing campaigns \citep{Coughlin19, Hosseinzadeh19, Lundquist19, Antier20, Gompertz2020}.  Each facility has different capabilities in aperture, field of view (FOV), and location.  Additionally, strategies related to choosing pointings, filters, and cadence resulted in a heterogeneous but vast dataset.  No candidate counterpart has been reported with high confidence in these data \citep[though see][for discussion of a low-significance fast radio burst (FRB) counterpart]{Moroianu23}, and the possible emission from a KN or sGRB has been limited by the multiple analyses from the above individual teams on their separate datasets. 

A combined analysis will clearly result in better constraints than analyses of subsets of the full dataset.  Here, we present our UVOIR search for an EM counterpart to GW190425 and combine those data with previously published data in Section~\ref{sec:data}.  We are left with 28 viable optical counterparts in this combined dataset, 4 of which were discovered within 5~days of merger, are confirmed in the localization volume, and have luminosities that could be consistent with KNe.

Our detailed candidate vetting process is described in Section~\ref{sec:candidates}. We introduce \teglon\footnote{\url{https://github.com/davecoulter/teglon_O4}}\textsuperscript{,}\footnote{\url{https://anathem.fandom.com/wiki/Teglon}} \citep[\coulterpub;][]{Coulter21}, a new, open-source tool for analyzing EM search data and performing pixel-level upper limits calculations in Section~\ref{sec:teglon}.
In Section \ref{sec:teglon_0425}, we describe how \teglon\ is used to perform a sophisticated analysis of the imaging data from all publicly reported observations and data newly reported here.  We account for the recovery of artificial sources in each image (when available), line-of-sight extinction, the three-dimensional (3D) probability from GW data, and available galaxy catalogs and their 3D completeness.  From this analysis, we present in Section~\ref{sec:model_comparison} the most comprehensive KN, sGRB, and model-independent constraints on the UVOIR emission from GW190425 under the assumption that no counterpart to GW190425 was found. Section~\ref{sec:discussion} discusses these results in the context of the LVK's current (i.e., fourth; O4) and future observing runs (O5+), and how future observational campaigns can adjust to improve our chances of discovering the next GW counterpart, along with a broader discussion of our analysis methods and prospects for improving the localizations of GW events based on contextual data. 

Throughout this paper, we adopt a flat $\Lambda$CDM cosmology with the following parameters: H$_{0}=100h=70$~km~s$^{-1}$~Mpc$^{-1}$, $\Omega_{\rm M}=0.27$, and $\Omega_{\rm \Lambda}=0.73$.

\section{Observations}\label{sec:data}


GW190425 \citep[denoted as S190425z by][]{S190425z-discovery} was initially reported to have a {\tt BAYESTAR} \citep{Singer16} 90\% credible localization of \initialninetyarea\ and a luminosity distance of $155 \pm 45$~Mpc. These were later refined to a final localization of \finalninetyarea\ and a
luminosity distance of 159$^{+71}_{-69}$~Mpc \citep[][]{Abbott20:gw190425}. Because of this large area, we consider any data across the sky relevant if obtained within two weeks of 25 April 2019, including targeted search data for GW190425 and GW190426\_152155, a purported NSBH merger with a final localization of 1,393~deg$^{2}$ and a luminosity distance of $377 \pm 45$~Mpc \citep[][]{O3a}. Therefore, we include in our analysis targeted search imaging data from the 1M2H Collaboration, from Gravity Collective (GC) partners Las Cumbres Observatory (LCO) and the 0.76\,m Katzman Automatic Imaging Telescope \citep[KAIT;][]{filippenko01} at Lick Observatory, from publicly reported limits through the Treasure Map (TM) application \citep[][]{Wyatt20}, and from the literature and private communication for the Pan-STARRS and ATLAS telescopes, for both GW190425 and S190426c. These TM data include limits for both GW events from the Zwicky Transient Facility \citep[ZTF;][]{Bellm2018}, the Gravitational-wave Optical Transient Observer 4 telescope \citep[GOTO-4;][]{GOTO_Survey}, the {\it Swift} Ultra-Violet/Optical Telescope \citep[][]{SwiftUVOT}, the MLS 10K CCD camera via the Catalina Sky Survey \citep[CSS;][]{CSS_Detector}, and the MMT Cam via the Fred Lawrence Whipple Observatory \citep[][]{MMT_Detector}. In addition to these targeted search data, we also include untargeted imaging data across all 1M2H telescopes, collecting a total of 3,598 public and private pointings for this analysis, which cumulatively cover 48.13\% of the two-dimensional (2D) probability and 48.28\% of \teglon-redistributed probability (see Section \ref{sec:teglon} and \ref{sec:teglon_0425}) from the final maps presented by \citet{Abbott20:gw190425}.

\begin{figure*}
    \centering
    \includegraphics[resolution=300,width=0.9\textwidth]{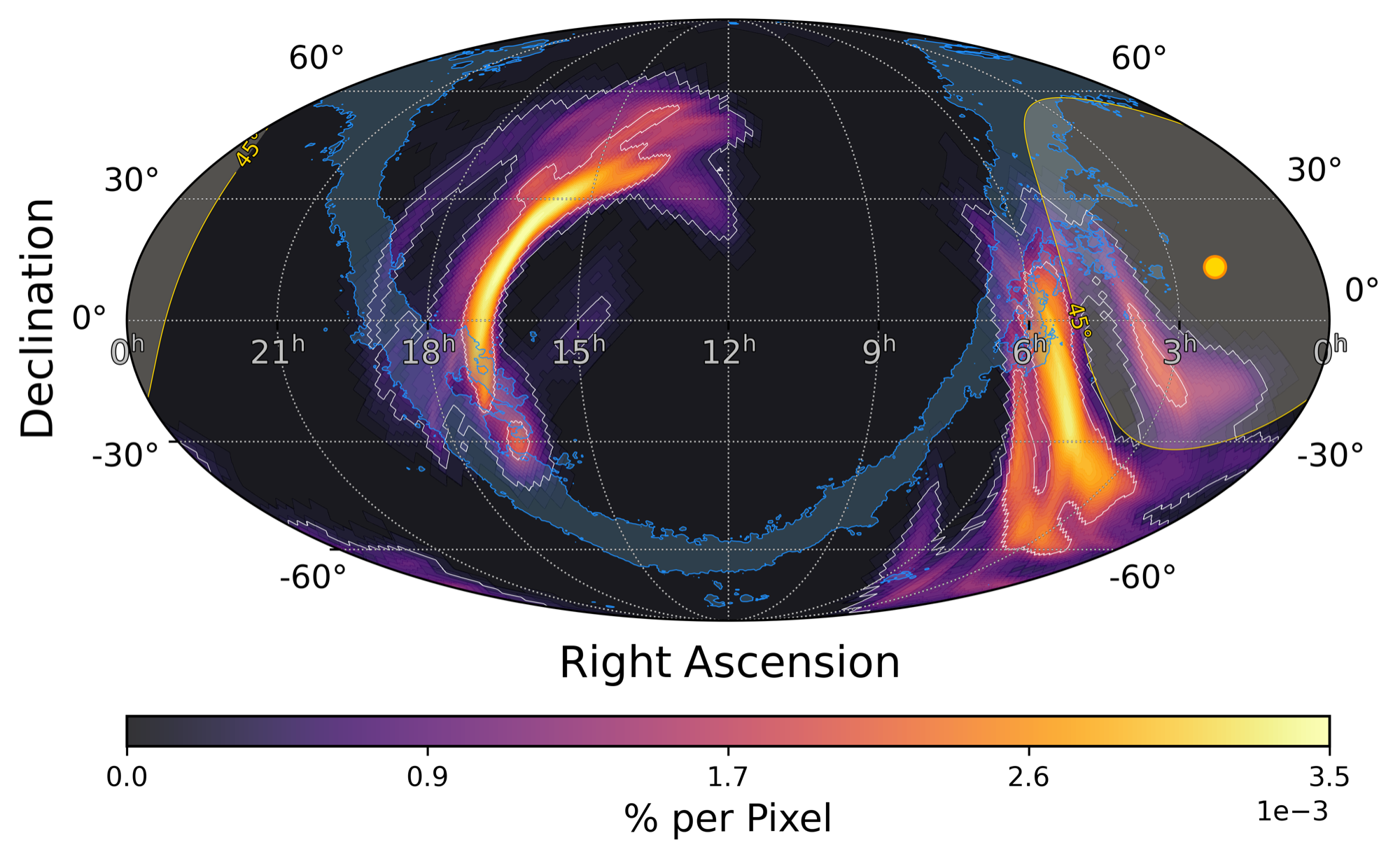}
    \caption{The LVC localization region for GW190425. Contours correspond to the 50th (\finalfiftyarea) and 90th (\finalninetyarea) percentile regions. In blue is the contour corresponding to the Milky Way $r$-band extinction of 0.5~mag. Near 3~hr of right ascension (R.A.) is the location of the Sun on 25 April 2019, with a yellow Sun-separation contour of 45$^{\circ}$.}
    \label{fig:2D_localization}
\end{figure*}

\subsection{One-Meter Two-Hemispheres Data}\label{sec:1m2h-data}

The 1M2H Collaboration was established in 2017 and originally used two 1\,m telescopes, the Nickel telescope at Lick Observatory in California and the Swope telescope at Las Campanas Observatory in Chile, to search for EM counterparts to GW sources. In 2019 this collaboration was expanded to include the 0.7\,m robotic Thacher telescope at the Thacher School Observatory in Ojai, CA \citep{Thacher22}, and the A Novel Dual Imaging Camera \citep[ANDICAM;][]{ANDICAM} on the SMARTS 1.3\,m telescope at Cerro Tololo Interamerican Observatory, Chile. We present data from this collaboration for the first time and describe our reduction process and limits below.

\subsubsection{ANDICAM}\label{sec:data_andicam}

We observed galaxies in the localization region of GW190425 with
ANDICAM.
All observations were performed from 25--26 April 2019 as described in Table~\ref{tab:observations}.  ANDICAM enables simultaneous optical observations using a charge-coupled device (CCD) with a $10' \times 10'$ FOV and IR observations using an array with a $3.3' \times 3.3'$ FOV.  We searched the initial localization with the CCD and IR detectors to obtain $I$- and $H$-band observations of 25 galaxies within the GW190425 90th percentile localization region and followed two optical candidates with a combined CCD + IR filter set of $I$, $J$, $H$, and $K$. All images for the CCD and IR detectors were reduced using {\tt photpipe} \citep{Rest05:photpipe}, including bias subtraction, dark corrections for the IR detector, and flatfielding.  The images were aligned using {\it Gaia} astrometric standards \citep{GaiaDR3}.  We then performed photometry in each image using {\tt DoPhot} \citep[][]{Schechter93}.  Finally, the images were flux calibrated using Pan-STARRS DR2 photometric standards \citep[][]{Flewelling2020} transformed into $I$-band \citep[following transformations by][]{Jester05} and 2MASS $H$-band standards \citep{Skrutskie06}. We obtained follow-up observations of each field to use as templates for subtraction from 5--11 June 2019.  After processing each image using the methods mentioned above, we subtracted the reference images from our science frames using {\tt HOTPANTS} \citep{Becker15}.  Final photometry for all transient sources in each difference image was obtained using a custom version of {\tt DoPhot}.

\subsubsection{Nickel}\label{sec:data_nickel}

We observed galaxies in the localization region of GW190425 with the Direct 2k $\times$ 2k ($6.8^{\prime} \times 6.8^{\prime}$) camera on the Nickel 1\,m telescope at Lick Observatory, Mt. Hamilton, California.  We performed targeted observations of candidate host galaxies in the $r$ band from 26 April 2019 to 9 May 2019, and we include in our analysis untargeted $BVri$ observations in the same date range that are also within the GW190425 90th percentile localization region.  All observations were reduced following the procedure described above for ANDICAM CCD imaging, including image subtraction with templates obtained from 22 April 2018 to 10 May 2020 and forced photometry on all candidate optical counterparts using {\tt DoPhot}.

\subsubsection{Thacher}\label{sec:data_thacher}

We observed GW190425 with the Andor 2k $\times$ 2k camera ($20.8^{\prime} \times 20.8^{\prime}$) on the 0.7\,m robotic Thacher telescope at the Thacher School Observatory in Ojai, CA \citep{Thacher22}.  We include $griz$ follow-up data targeting the 90th percentile localization region of GW190425 obtained from 26 April 2019 to 4 May 2019.  All imaging was reduced following the aforementioned methods and by \citet{Kilpatrick21}.  Template imaging of each field was obtained from 23 February 2019 to 6 August 2021.  Our final observation list is given in Table~\ref{tab:observations}.

\subsubsection{Swope}\label{sec:data_swope}

We observed the localization region of GW190425 with the Direct 4k $\times$ 4k camera ($29.8^{\prime}\times29.7^{\prime}$) on the 1\,m Swope telescope at Las Campanas Observatory, Chile.   Our Swope observations consisted of targeted observations within the 90th percentile localization region of GW190425 in the $i$ band obtained from 25 April 2019 to 9 May 2019 and untargeted $uBVgri$ observations within the same area and time frame.  These data were reduced following the ANDICAM/CCD procedures described above.  We obtained template imaging from 16 August 2018 to 25 February 2020 to perform image subtraction in each frame and search for optical transients, and to generate forced photometry on known optical transients in each image.

\subsubsection{SN~2019ebq and MOSFIRE}\label{sec:candidates_2019ebq}

\begin{figure} 
    \includegraphics[width=0.49\textwidth]{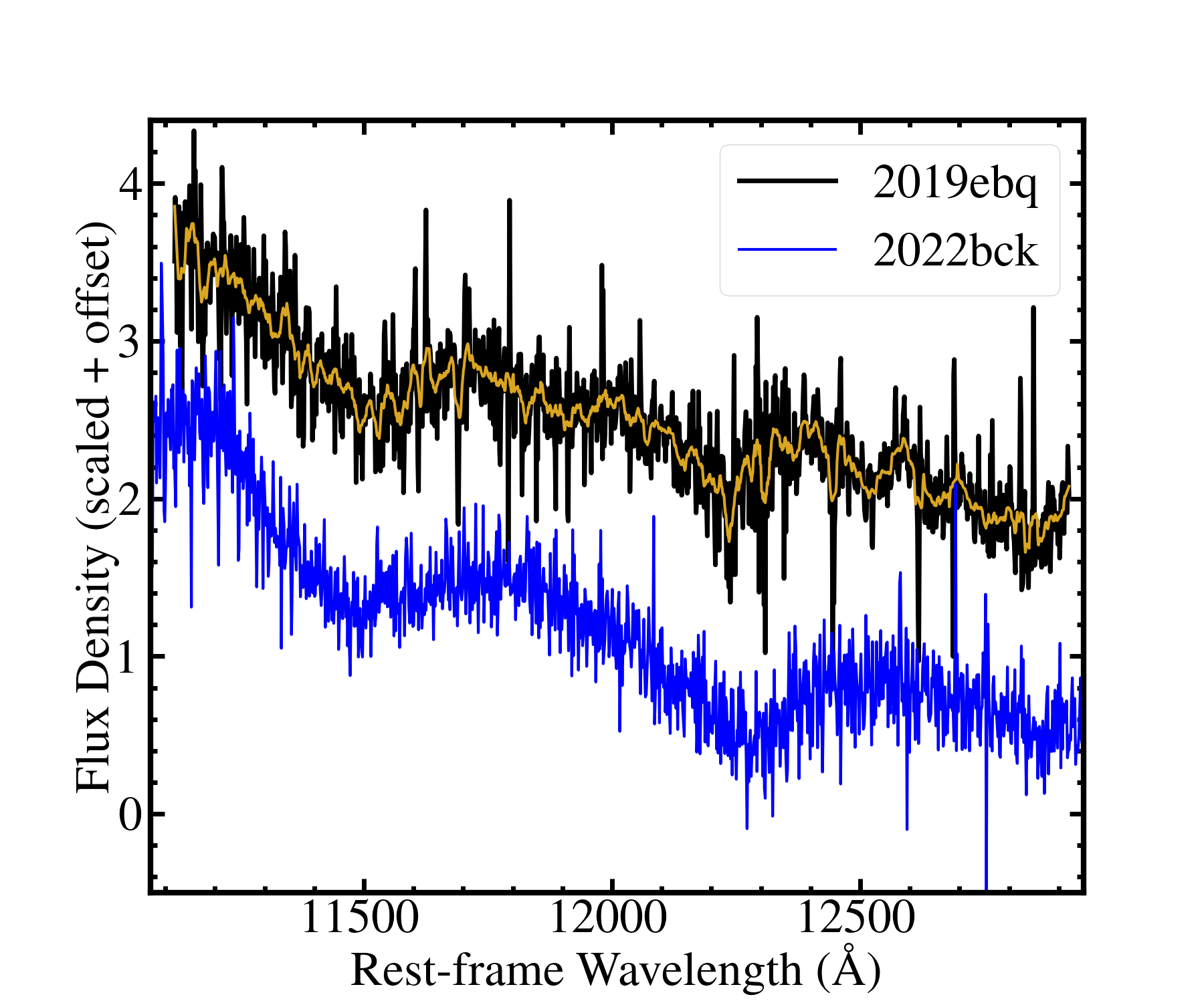}
    \caption{MOSFIRE NIR spectrum of SN\,2019ebq (black), covering the $J$ band blueward of 12,900~\AA, obtained on 2019 Apr 26, 14:32:11 UTC on Keck I.  The spectrum has been smoothed to a resolution of 17~\AA\ (yellow) for analysis of the broad transient features present at these wavelengths. Our SN\,2019ebq MOSFIRE spectrum appears consistent with a Type Ib/c SN, and is therefore unrelated to the GW event \citep[see \ref{sec:candidates_2019ebq};][]{2019ebq_GCN}.  We demonstrate this point by comparing it to a Keck/NIRES spectrum of the Type Ib SN\,2022bck presented by \citet{Tinyanont22}.  The similarity between these spectra reinforces the nature of this object as an SN and not a GW counterpart.}
    \label{fig:2019ebq}
\end{figure}

We obtained near-infrared (NIR) spectroscopy of the candidate counterpart to GW190425 SN\,2019ebq on 2019 Apr 26, 14:32:11 UTC with the Multi-Object Spectrometer For Infra-Red Exploration \citep[MOSFIRE;][]{mosfire} on the Keck I 10\,m telescope.  The spectrum was originally reported and described by \citet{2019ebq_GCN}.  We reduced the spectrum following standard procedures using {\tt spextool} and show the final reduced spectrum in Figure~\ref{fig:2019ebq}.  Similar to findings by \citet{2019ebq_GCN} and \citet{2019TNSCR.642....1N}, we classify this source as an SN\,Ib/c near peak light based on comparison to a Keck/NIRES spectrum of the Type Ib SN\,2022bck presented by \citet{Tinyanont22}.


\subsection{Gravity Collective Data}\label{sec:gwc-data}

\subsubsection{KAIT}\label{sec:data_kait}

The 0.76\,m KAIT \citep[][]{Richmond93, Filippenko01} at Lick Observatory targeted galaxies in the localization regions of GW190425 and GW190426 between 2019 April 25 and 27, as described by \citep[][]{KAIT0425GCN, KAIT0426GCN}. Galaxies were selected from GLADE \citep[][]{Dalya18}, according to their $B$-band luminosity, with target priority reweighted by elevation at the time of observation. All observations were performed in a ``Clear'' filter. 688 galaxies were targeted between both events, with all fields being reimaged in July 2023 to provide templates of the same fields for detailed analysis. Following standard imaging and photometry procedures (e.g., \citep[][]{Ganeshalingam10}; \citep[][]{Zheng18}), the images were calibrated, and point-spread-function (PSF) photometry was performed using DAOPHOT \citep{Stetson87} in {\tt IDL}. The throughput of the KAIT ``Clear'' filter is close to that of the $R$ band \citep{Li03}, so local AAVSO Photometric All-Sky Survey (APASS) standards \citep{Henden15} were transformed to the \citet{Landolt92} $R$ band following \citet{Jester05}. Template images were then subtracted from the August 15 and 18 epochs using a custom {\tt IDL}-based image-subtraction pipeline for PSF convolution. Finally, we estimate the limiting magnitude in each subtracted image using the flux-weighted average of the sky background in the convolved science and template frames, which is reported in Table~\ref{tab:observations}.

\subsubsection{LCO}\label{sec:data_lcogt}

The Gravity Collective combines follow-up efforts by 1M2H and the Las Cumbres Observatory (LCO) Global Telescope network \citep{Brown13}, which includes fourteen 0.7-1\,m telescopes distributed worldwide. LCO observed the localization region of both GW190814 and S190426c, with a galaxy-targeted search and prioritization strategy described by \citep[][]{Arcavi17a}. For both GW events, LCO obtained 773 exposures of duration 300~s each in $gri$ using the Sinistro cameras ($26^{\prime}\times26^{\prime}$ FOV) mounted on these telescopes (\lcogtpub). Image processing was performed by the LCO {\tt BANZAI} pipeline \citep[][]{McCully18}, and limiting magnitudes were calculated using {\tt LCOGTSNpipe} \citep{Valenti16}. Sloan Digital Sky Survey (SDSS) \citep{SDSSDR7}, PS1 \citep{Flewelling2020}, or DECam \citep{DESDR1} template images were used in the science image bands to perform image subtraction using {\tt PyZOGY} \citep{Zackay16, Guevel17}. The limiting magnitudes were calculated by determining the Poisson noise due to the sky using the median absolute deviation of the entire image. The Poisson and read noise were combined, and the 3$\sigma$ limiting magnitude (median limiting magnitude of 22.1 mag) was estimated by inverting the standard signal-to-noise ratio (S/N) equation.

\subsection{Public Data via Treasure Map}\label{sec:public-data-tm}

\subsubsection{ZTF}\label{sec:data_ztf}

ZTF is a 47 deg$^{2}$ FOV optical instrument on the Palomar 48-inch Schmidt telescope \citep[][]{Bellm2018}. We include 313 ZTF pointings reported to TM with a status of ``completed'' for GW events GW190425 and GW190426\_152155, and whose image-reduction process is outlined by \citep[][]{Coughlin19}. Pointings span the $g$, $r$, and $i$ bands, with a median $r$-band depth of $\sim 21.5$ $m_{\mathrm{AB}}$. Within TM, each pointing includes the central coordinate of the FOV, filter, MJD of the observation, and limiting AB magnitude.


\subsubsection{CSS}\label{sec:data_css}

CSS operates the MLS 10k CCD camera on the Mt. Lemmon 1.5\,m telescope, which has a $\sim 5$ deg$^2$ FOV and was used by the Searches after Gravitational Waves Using ARizona Observatories (SAGUARO) team to search for 17 GW events within O3 \citep[][]{Lundquist19, Paterson20}. We include 61 pointings taken in an open filter to a median limiting mag of $\sim 21.3$ $m_{\mathrm{AB}}$.


\subsubsection{GOTO-4}\label{sec:data_goto}

The GOTO-4 telescope \citep[][]{GOTO_Survey} is a prototype array of 4 telescopes with a combined FOV of $\sim 18$ deg$^{2}$. The GOTO team searched for 29 GW event triggers in LIGO's O3 \citep[][]{Gompertz2020}, and we include 399 pointings that span the $g$ and $V$ bands, with a median $g$-band depth of $\sim 19.8$~$m_{\mathrm{AB}}$.

\subsubsection{MMT}\label{sec:data_mmtcam}

The 6.5\,m MMT at Fred L. Whipple Observatory in Arizona conducted a galaxy-targeted search with MMTCam for EM counterparts to both GW190425 and S190425c \citep{Hosseinzadeh19}. We include 119 pointings in $g,i$, with a median $i$-band depth of 21.9~$m_{\mathrm{AB}}$.

\subsubsection{Swift}\label{sec:data_swift}
In O3, {\it Swift} searched 18 GW events using a galaxy-targeted approach \citep{Evans16}, including GW190425 and S190426c, with the Ultra-Violet/Optical Telescope \citep{Oates21}. These data include 1357 pointings for GW190425 and S190426c in the $u$ band with a median limiting magnitude of 19.4~$m_{\mathrm{AB}}$. These data are particularly interesting as they cover a region of parameter space which is unique given the other optical filters in this combined dataset. 

\subsection{Public Data via Literature Review}\label{sec:public-data-lit}

\subsubsection{Pan-STARRS}\label{sec:data_PS1}
The Pan-STARRS data used in this study come from the Pan-STARRS 1 (PS1) telescope, at the summit of Haleakala on the Hawaiian island of Maui. PS1 is equipped with a composite 1.4 Gigapixel camera and has a FOV of 7.06 deg$^2$ in a circular aperture; this FOV is broken into a tessellation of ``skycells''\footnote{\url{https://outerspace.stsci.edu/display/PANSTARRS/PS1+Sky+tessellation+patterns}}, each having a dimension of $\sim 24' \times 24'$ \citep{Chambers2016}. Each skycell can be treated as an individual pointing that inherits the parent's world coordinate system and image depth, and allows PS1 upper-limits data to be more easily analyzed by \teglon. The PS1 telescope began searching the localization of GW190425 $\sim 80$~min after the GW trigger, and continued for the following three days, publishing their upper limits \citep{Smartt24}. The pointings and limits derived from these data were shared with our team via private communication and include 6,558 skycells covering a unique sky area of $\sim$\panstarrsarea in the $i$ band to an average depth of $\sim 21.5$~mag.

\subsubsection{ATLAS}\label{sec:data_atlas}
The Asteroid Terrestrial-impact Last Alert System (ATLAS) telescope system is a network of four identical telescopes, with two telescopes located on the Hawaiian Islands (one on Haleakala and another on Mauna Loa), one telescope in El Sauce, Chile, and the last telescope in Sutherland, South Africa \citep{Tonry2018}. Each telescope has a rectangular FOV of 28.9~deg$^2$, and over the course of $\sim7$ days observed the localization of GW190425, publishing their upper limits \citep{Smartt24}. The pointings and limits derived from these data were shared with our team via private communication and include 437 pointings covering a unique sky area of $\sim$\atlasarea in the ATLAS $o$ and $c$ bands to an average depth of $\sim18.9$~mag.


\begin{figure*}
    \centering
    \includegraphics[resolution=300,width=0.9\textwidth]{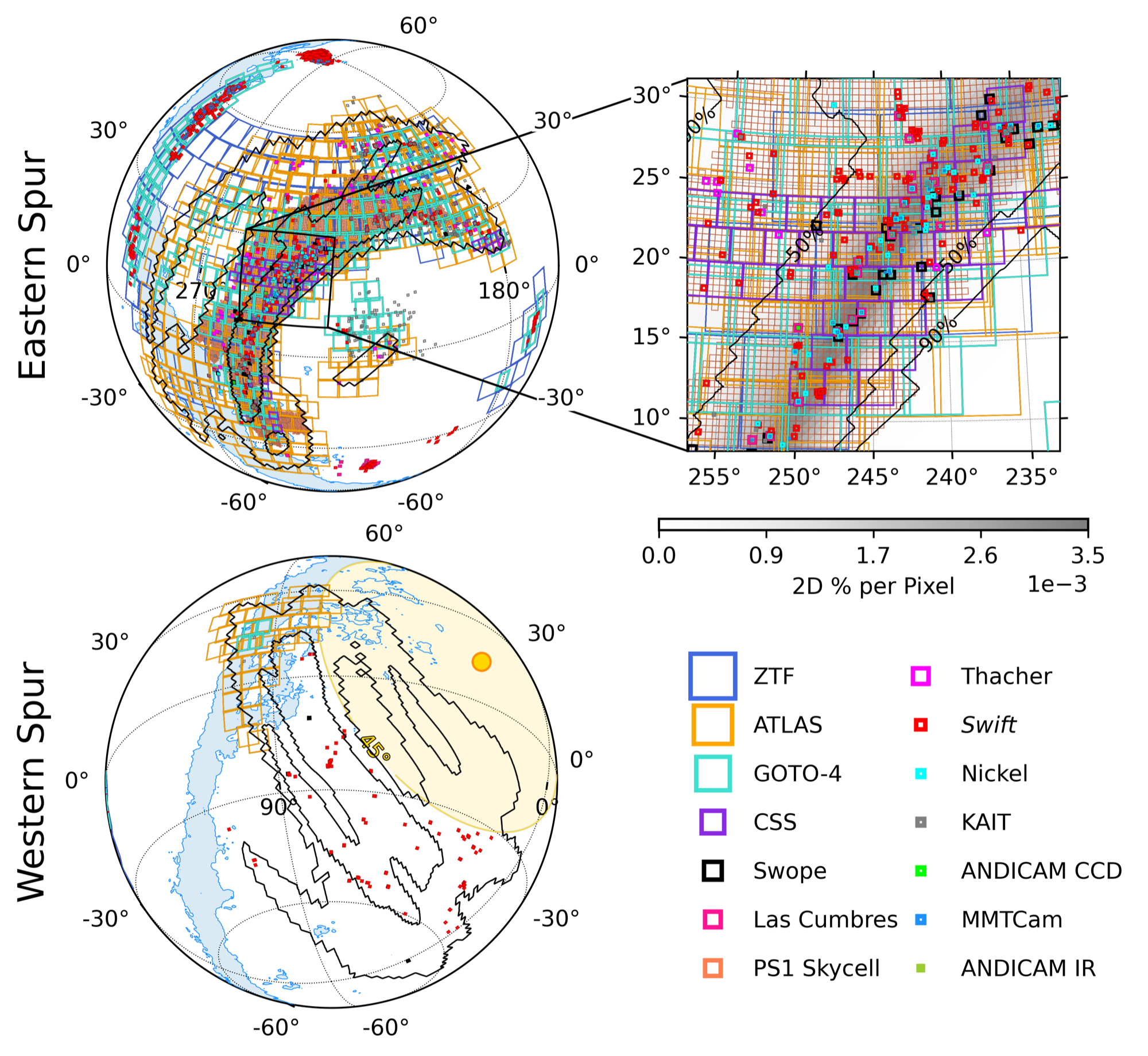}
    \caption{Visualization of the full EM search dataset for this work. In each orthographic projection, black contours correspond to the 50th/90th localization of GW190425, and Milky Way extinction is marked as a blue contour at ${\rm A}_{r}=0.5$~mag. Overplotted are all instrument FOVs: ZTF (dark blue), ATLAS (light orange), GOTO-4 (turquoise), CSS (purple), Swope (black), Las Cumbres (magenta), PS1 Skycells (dark orange), Thacher (hot pink), {\it Swift} (red), Nickel (cyan), KAIT (gray), ANDICAM CCD (lime green), MMTCam (light blue), and ANDICAM IR (dark green). {\it Upper left:} Close-up view of the Eastern Spur of probability as seen in Figure \ref{fig:2D_localization}, R.A. $\in[95, 315]$~deg. Every instrument in our manifest has imaging in this hemisphere (10,365 total field centers; 266 not shown), covering 45.2\% of the final LVC 2D probability. {\it Upper right:} A $12^\circ$-radius zoom-in on R.A. 245 deg, Declination (Decl.) $+$20 deg, showing a detailed view of the smaller FOV instruments. The grayscale gradient is the 2D probability of the localization, with the 50th/90th contours labeled. Many of these fields covered the same sky regions multiple times in the same filters and highlight an opportunity to coordinate EM follow-up efforts (see Section \ref{sec:discussion}).  {\it Lower left:} Close-up view of the Western Spur of probability as seen in Figure \ref{fig:2D_localization}, R.A. $\in[0, 95]$~deg. Only ATLAS, GOTO-4, Swope, and {\it Swift} have observations in this hemisphere (355 total) owing to this region being close to the Sun, contributing only 3.1\% of the covered 2D probability. A yellow Sun contour denotes a $45^\circ$ separation that marks {\it Swift}'s pointing limits.}
    \label{fig:survey_footprint}
\end{figure*}

\section{Candidates}\label{sec:candidates}
\subsection{1M2H Vetted Candidates}\label{sec:candidates_sniffed}

After subtracting templates from the ANDICAM, Nickel, Swope, and Thacher images, we identified candidate counterparts to GW190425 by searching for sources of positive emission in the difference images.  We first ran {\tt DoPhot} on the difference images, searching for sources detected at a S/N threshold of $\geq 3\sigma$.  We performed minimal filtering on the detected sources, particularly removing those where $>30$\% of pixels inside the PSF aperture are negative or where $>40$\% of pixels are masked.  Apart from these cuts, we required only that a candidate transient is detected in a single image at our S/N threshold.

All candidates were then gathered by field into web pages with cutout images showing the candidate detection from every epoch, the scatter in candidate coordinates for each detection, and the difference light curve in terms of flux and magnitude.  Members of the 1M2H collaboration all visually vet these web pages to rule out detections that appear consistent with artifacts such as a convolved cosmic ray, correlated noise across a bad section of each detector, dipole emission associated with a bright and poorly subtracted star, or a satellite or other moving object passing through the image frame.

\begin{figure}
    \vspace*{0.8em}
    \includegraphics[width=0.49\textwidth]{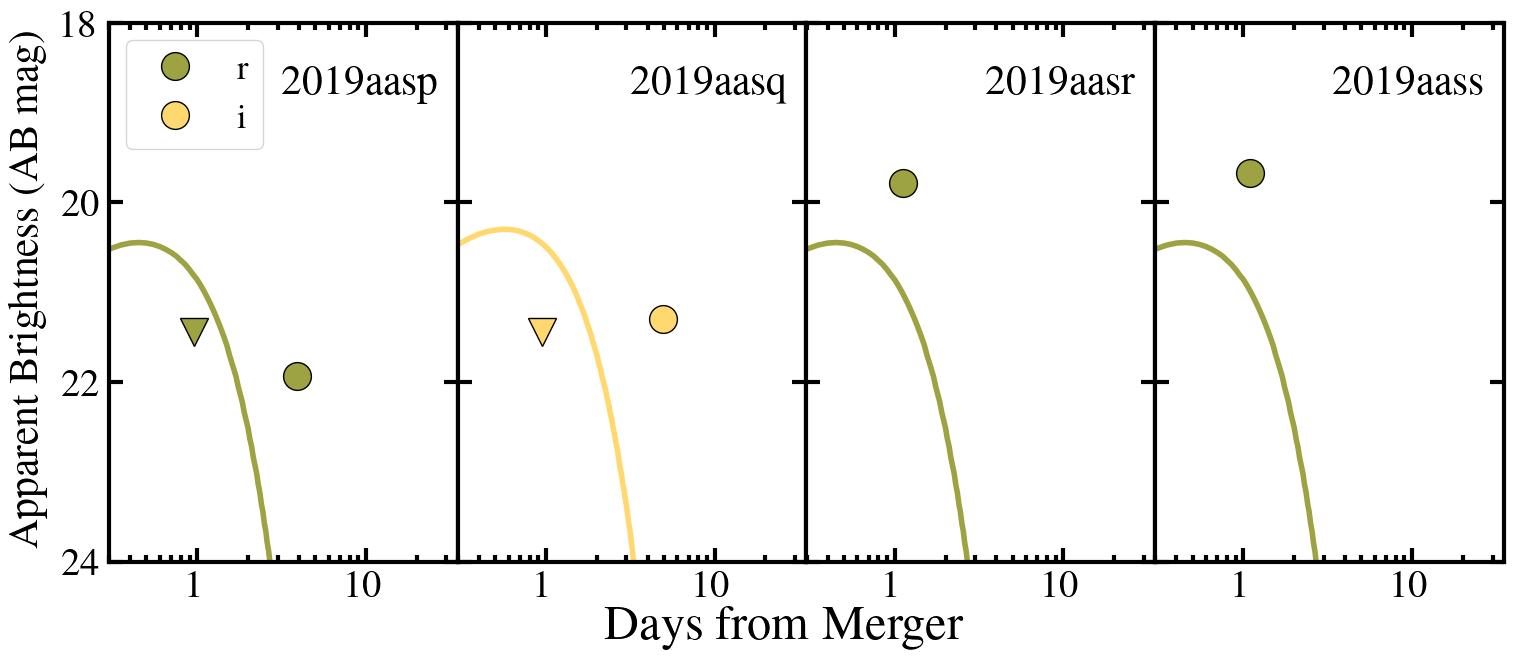}
    \caption{Photometry from the four candidate counterparts to GW190425 discovered within the localization region by 1M2H and described in Section~\ref{sec:candidates_sniffed}.  We show the time of detection as a circle in each panel, with green corresponding to $r$-band photometry and yellow to $i$-band photometry.  For comparison, we overplot model KN light curves for a hypothetical event with ejecta mass 0.023\,M$_{\odot}$, velocity 0.26$c$, and an electron fraction $Y_{e}=0.45$ as described in Section~\ref{sec:model_comparison_kne}.}
    \label{fig:cand_lightcurves}
\end{figure}

We required that a candidate transient be flagged only by a single human vetter to elevate that source for our candidate analysis pipeline.  Following analysis similar to that of \citet{Kilpatrick21} and public candidates described below, we crossmatched the candidates to sources classified as stars by {\it Gaia} \citep[point source score (PSS) $>0.99$ following the PSS value from][]{GaiaDR3}, were within 2\arcsec\ the location of a minor planet at the time of observation based on ephemeris from the Minor Planet Catalog\footnote{\url{https://minorplanetcenter.net}}, or were crossmatched to known, public transients in the Transient Name Server\footnote{\url{https://www.wis-tns.org/}}.  After these checks, we identified four novel candidate transients that were reported to TNS: AT~2019aasp, 2019aasq, 2019aasr, and 2019aass \citep{1M2H-transients}.  These and all other candidate transients reported to TNS were then analyzed using methods described below.  In Figure~\ref{fig:cand_lightcurves} we show photometry from candidates discovered by 1M2H in comparison to model KN light curves described in Section~\ref{sec:model_comparison_kne}.

\begin{figure}
    \hspace*{-2.2em}
    \includegraphics[width=0.52\textwidth]{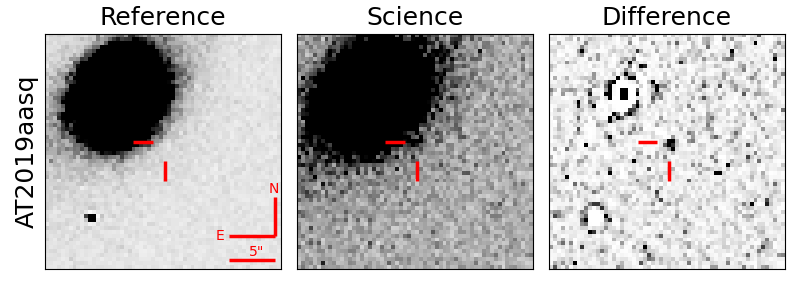}
    \caption{Image cutout triplet for AT~2019aasq, a counterpart candidate discovered by the Swope telescope +4.93~days after the GW trigger, that our analysis cannot rule out as a candidate counterpart to GW190425 and which we characterize as a ``more likely'' kilonova candidate (see Sections \ref{sec:candidates} and Appendix \ref{sec:candidate_counterparts} for a detailed discussion). Both the reference ({\it left}; MJD 58638.18288) and science ({\it middle}; MJD 58603.27694) images were obtained by Swope and show the same $26.1^{\prime\prime}\times26.1^{\prime\prime}$ region of sky centered on AT~2019aasq.  The difference image ({\it right}) highlights the discovery detection of the transient at $i=21.30\pm0.19$~mag. We note that the candidate is clearly visible $\sim 7.5$\arcsec\ from the center of its likely host galaxy WISEA J154032.14+282013.7 at $z=0.031090$ \citep{2023TNSTR1610....1C}, within the 1$\sigma$ most likely GW190425 volume. AT~2019aasq was not detected with either ATLAS $c$ or $o$ bands, or ZTF $g$ or $r$ bands, despite coverage within days of this detection, implying it was likely an intrinsically red, faint transient. }
    \label{fig:cand_cutout}
\end{figure}

\subsection{Public Candidates}\label{sec:candidates_public}

We used our candidate analysis pipeline to vet candidate counterparts to GW190425 in the context of the final {\tt GW190425\_PublicationSamples}\footnote{\url{https://gracedb.ligo.org/superevents/S190425z/files/}} localization map, time discovered from merger, coincidence with likely stars or other known point sources and minor planets, spectral classification as a transient type unlikely to be associated with an NS merger, association with a host galaxy outside the localization volume defined by the {\tt bayestar} map, and photometric evolution that does not resemble a likely KN or afterglow counterpart.  In general, these cuts follow the methods described by \citet{Kilpatrick21} and the examples implemented by \citet{CAP}.  Here we summarize each step.

\begin{enumerate}
    \item Importation of candidates from our transient database {\tt YSE-PZ} \citep{Coulter23}, which includes all transients and metadata contained in TNS.
    \item We analyze only candidates discovered within the first 14~days after the coalescence time of GW190425 on 2019 April 25, 08:18:05 UTC as defined by \citet{Abbott20:gw190425}.  Moreover, we only analyze candidates within the 2D 90th percentile as defined by the final {\tt GW190425\_PublicationSamples} map of that event.  These two initial cuts define our sample of \Ncandidate\ candidate counterparts analyzed in the remaining steps below.
    \item We crossmatch to minor planets using the time of discovery and coordinates of each candidate and using the Minor Planet Checker\footnote{\url{https://cgi.minorplanetcenter.net/cgi-bin/checkmp.cgi}}.  Any source found within 2\arcsec\ of a known minor planet at the time of observation is ruled out.  In total, \Nmpc\ candidates were ruled out by this check.
    \item We crossmatch to point sources within the {\it Gaia} \citep{GaiaDR3} and Pan-STARRS DR2 catalogs \citep{Flewelling2020}.  For {\it Gaia}, this involves checking for sources aligned within 2\arcsec\ of a source with point-source score $>0.99$, while for Pan-STARRS we check for candidates within 2\arcsec\ of a source classified as point-like by the PS1 detection-flagging algorithm\footnote{\url{https://outerspace.stsci.edu/display/PANSTARRS/PS1+Detection+Flags}}.  \Ngaia\ candidates were ruled out for coincidence with {\it Gaia} sources while no candidates were ruled out owing to coincidence with Pan-STARRS sources.
    \item For candidates with spectroscopic follow-up observations, we rule out those with a spectral classification that is inconsistent with a KN or GRB afterglow.  For GW190425, this sample comprises sources classified in TNS as a cataclysmic variable (CV), SLSN, SN\,Ia, SN\,Ib, SN\,Ic, SN\,II, and SN\,IIn, which are known to arise from progenitor systems other than NS mergers.  \Nclass\ candidates were ruled out based on their spectral classifications.
    \item We rule out candidates with pre-merger activity within 2\arcsec\ of the transient location based on a positive detection using forced photometry in ASAS-SN \citep{Kochanek17}, ATLAS \citep{Tonry2018, Smith2020, Shingles2021}, or ZTF \citep{Bellm2018}.  For additional details on our querying method, see \citet{Coulter23}.  We rule out \Nvariable\ candidates that have premerger variability.
    \item In addition to candidates outside the nominal localization area, we associate all candidates with host galaxies when possible and rule out candidates that are outside the 90th percentile localization volume defined by the final {\tt GW190425\_PublicationSamples} localization map.  We derive our host-galaxy sample from those with spectroscopic redshifts in the NASA/IPAC Extragalactic Database (NED)\footnote{\url{http://ned.ipac.caltech.edu/}} or a photometric redshift from the PS1-STRM \citep{PS1redshift}, Photometric Redshifts for the Legacy Surveys \citep[Legacy;][]{Zhou20}, or 2MASS Photometric Redshift \citep[2MRS;][]{Bilicki13} catalogs.  Note that we place priors on the galactocentric host offsets of $<300$\arcsec\ and $<75$~kpc \citep[consistent with the maximum short GRB host offsets identified by, e.g.,][]{Fong13,Fong22} in selecting the most likely host galaxy, then associate each transient with the galaxy that provides the smallest physical offset from the GW candidate.  In this way, we ruled out \Nredshift\ candidates.
    \item Finally, we rule out candidates with photometry whose absolute magnitude, decline rate, or color evolution appears inconsistent with KN or afterglow emission.  The details of this calculation are described by \citet{Kilpatrick21}.  At this stage, there remained \Nbeforephot\ viable candidates, of which we ruled out \Nphot\ owing to photometric evolution inconsistent with being a counterpart to GW190425.  There remain \Nremain\ viable candidate counterparts.
\end{enumerate}

Based on each of the steps described above and the lack of spectroscopic follow-up observations, we cannot definitively characterize any of these sources as a GW counterpart (i.e., as a KN or GRB afterglow), so additional analysis of each source has limited utility.  However, taking the remaining sources in our analysis, we can differentiate between sources that are more likely to be GW counterparts versus interloping transients.  We provide a more detailed description of four of the remaining candidates considered to be the most likely counterparts to GW190425 in Appendix~\ref{app:candidates}.  

\begin{figure*}
    \centering
    \includegraphics[resolution=300,width=0.9\textwidth]{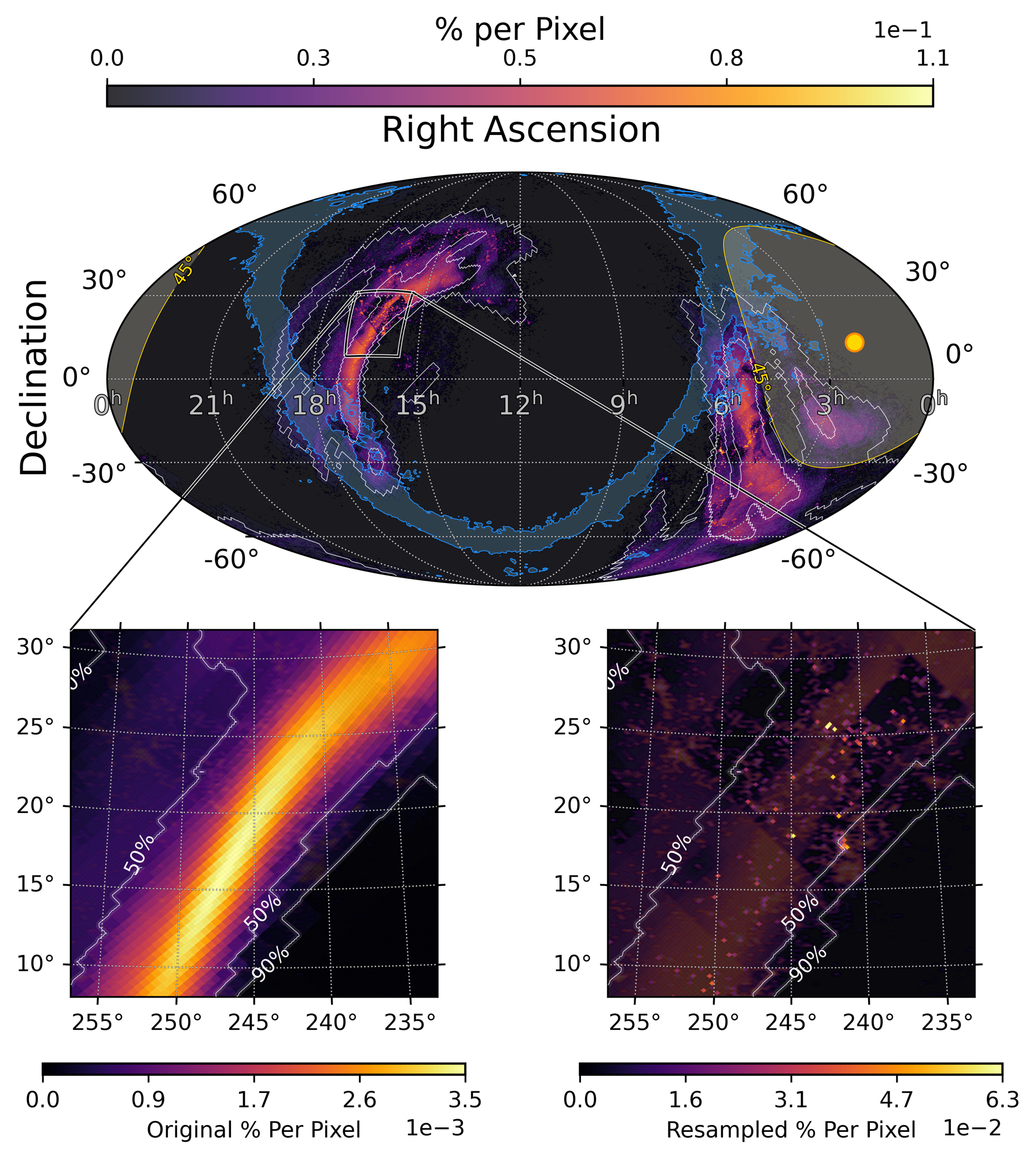}
    \caption{{\it Top:} The localization map resampled by \teglon. At GW190425's distance, \teglon\ redistributes half of the total 2D probability to the highest probability galaxies (see Section \ref{sec:teglon_0425}). A $12^\circ$-radius zoom-in panel is marked by a square centered on R.A. 245 deg, Decl. $+$20 deg. All other attributes are the same as in Figure \ref{fig:2D_localization}. {\it Bottom left:} For comparison, we show a zoom-in of the original localization map with white contours denoting the original 50th/90th localization. Within the bounding box of R.A. $\in[235, 255]$ deg and Decl. $\in[10,30]$ deg, there is $\sim 8.2$\% of the localization probability within $\sim 377$~deg$^2$. {\it Bottom right:} The same zoom-in region with \teglon's redistribution algorithm (matching the top plot). The same amount of probability ($\sim 8.2$\%) is covered in only $\sim 100$~deg$^{2}$, increasing the coverage efficiency by a factor of $\sim3.8$.}
    \label{fig:redistributed_localization}
\end{figure*}

\section{\teglon}\label{sec:teglon}

One effective method for localizing the EM counterparts of GW sources is to target the bright/massive galaxies residing in the locus of high probability within a GW localization volume \citep{Kanner12, Gehrels16}. This technique relies on two key factors: a galaxy catalog that is relatively complete at the ranges where GW sources are likely to be detected, and localization regions that are small enough to be efficiently searched with ground-based telescopes. To date, several catalogs have been used in GW follow-up searches \citep{Kopparapu08, White11, Dalya18, Dalya22, Cook23}, and all contain the key attributes of position, distance, and $B$-band magnitudes. The $B$ band is used in particular because the rate of BNS mergers is expected to follow the star-formation rate (SFR) in the local universe, and $B$ is a convolution of this SFR with a galaxy's total stellar mass \citep{Phinney91, Belczynski02}. In 2017, this technique led to the discovery of the first optical counterpart, AT~2017gfo \citep[][]{Coulter17}, to a GW source, GW170817. GW170817 was the first-ever BNS merger detected in GWs, and localized to an area of 31~deg$^{2}$ and a luminosity distance of $40^{+8}_{-14}$~Mpc \citep{Abbott17:detection}. In searching for this counterpart, 1M2H used the Gravitational Wave Galaxy Catalogue (GWGC), which at 40~Mpc is nearly 100\% complete when compared to a Schechter galaxy luminosity function \citep{Schechter76} for galaxies with a characteristic luminosity of $\leq -20.3$~mag.

However, as the LVK has improved the GW network detection sensitivity \citep{Abbott:Local}, these catalogs and techniques have become less effective. In O3, the typical BNS inspiral range was $\sim 108$--135~Mpc \citep{O3a}. For BNS mergers $>2.8$~M$_{\odot}$, or face-on mergers, the detection distance may be much larger. To this point, GW190425 was a uniquely massive BNS merger at 3.4$^{+0.3}_{-0.1}$~M$_{\odot}$, and was detected at a luminosity distance of $159^{+69}_{-71}$~Mpc \citep{Abbott20:gw190425}. At this distance, the Galaxy List for the Advanced Detector Era \citep[GLADE;][]{Dalya18} catalog is only $\sim 50$\% complete in galaxy luminosity, meaning that a naive approach of simply targeting bright galaxies in the catalog would miss half the total galaxy luminosity in the volume. Despite this, as BNS detection ranges increase, the surface density of galaxies in projection increases so that any field of view should contain many galaxies (both cataloged and uncataloged).  Naively, a pure tiling approach to searching for a counterpart is more effective at larger distances, but this picture is complicated by inhomogeneous galaxy catalog coverage. Intelligently trading off between these two approaches --- to use our knowledge of where galaxies are to target them and to tile the localization region when we do not --- motivates the creation of a new tool called \teglon. A detailed treatment of how \teglon\ transforms GLADE, implements its completeness weighting, and calculates its pixel-level upper limits will be presented in a forthcoming companion paper.

\subsection{GW190425 Transformed By \teglon}\label{sec:teglon_0425}

\teglon's EM search optimization depends on two properties: the $B$-band luminosity completeness of its volume pixels, or voxels, in the GW localization volume, and how much area that GW localization volume subtends on the sky in projection. Completeness is largely dictated by the average luminosity distance to an event; however, in regions of high galaxy catalog completeness (e.g., SDSS Stripe 82 \citep{Annis14} or the survey footprint of 2dF \citep{Colless01}), \teglon's algorithm can still be effective at redistributing probability in the original GW localization to high-probability galaxies, thereby reducing the area an instrument needs to search. However, the area in projection of a GW localization also matters --- if the area subtended by a GW localization fits within the FOV of a search instrument (e.g., GW170817's localization ), redistributing probability on scales smaller than the FOV would not change the search strategy. In the edge case of 0\% completeness, or very small projected areas, \teglon's~optimization is identical to a pure tiling pattern of the high-probability region.

The localization of GW190425 is a quintessential use case for \teglon. Because this event was only detected by the LIGO Livingston detector, its location was constrained to nearly a quarter of the sky (\finalninetyarea). However, despite this large area, the distance was relatively close at $159$~Mpc \citep{Abbott20:gw190425}. At this distance, the GLADE catalog is on average $\sim 50$\% complete, and therefore \teglon\ redistributed half of the localization probability to galaxies at the correct distance. This resampling reduces the 90th percentile localization to \fourDninetyarea, a factor of $\sim1.5$. Figure~\ref{fig:redistributed_localization} shows the resulting localization, with insets that highlight this updated concentration of probability.

Table~\ref{tab:search_synopsis} shows a synoptic view of the effect \teglon\ has on the search efficiency increase, $\eta$, for each instrument in the dataset. The value of $\eta$ is markedly enhanced for detectors with FOVs $\le 1$~deg$^{2}$. In general, these instruments followed a galaxy-targeted approach, and owing to the high completeness of GLADE with respect to GW190425's localization, the redistributed map provided by \teglon\ confirmed that these galaxies were in regions of the sky more likely to host the progenitor of the GW event. For this particular event, 3,402 of the original map pixels ($\sim 178$~deg$^{2}$) had their probability values boosted by factors of $\ge10$ over their original values, and constitute 16\% of the total probability in the map.

For this dataset, while all instruments have $\eta\ge1.0$, instruments with FOVs $\ge 1$~deg$^{2}$ saw diminishing returns owing to the fact that their large footprint on the sky allowed them to simply tile the entire Western Spur of the localization (see Figure \ref{fig:survey_footprint}). Because of this, the survey footprint of these instruments encompassed both the pixels where probability was being concentrated and the voids left in between, resulting in $\eta$ approaching unity. However, in the maximal case where a GW event subtends the entire sky but is detected at a distance where GLADE is 100\% complete (e.g., at the distance of GW170817; 40~Mpc), \teglon\ would be useful for even the largest FOV instruments.

\begin{deluxetable*}
            {lccccc}
            \tabletypesize{\scriptsize}
            \tablecaption{GW190425 Search Synopsis\label{tab:search_synopsis}}
            \tablewidth{0pt}
            \tablehead{
            \colhead{Search Instrument} & 
            \colhead{FOV} & 
            \colhead{\# of Fields} & 
            \colhead{Total 2D Probability} & 
            \colhead{Total Redistributed Probability\tablenotemark{\footnotesize a}} & 
            \colhead{Efficiency Increase} \\ [-0.1cm]
            &
            (deg$^{2}$) &
            & 
            $\sum_{i}$$\ptDi$(\%) &
            $\sum_{i}$$\ppptDi$(\%) & 
            $\eta$$\equiv$$\sum_{i}$$\frac{\ppptDi}{\ptDi}$
            \\ [-0.45cm]
            }
\startdata
ANDICAM IR & 0.0015 & 21 & 0.04 & 0.46 & 11.36 \\
MMTCam & 0.0020 & 118 & 0.15 & 1 & 6.44 \\
ANDICAM CCD & 0.0112 & 27 & 0.06 & 0.62 & 10.39 \\
KAIT & 0.0128 & 412 & 0.23 & 1.7 & 7.37 \\
Nickel & 0.0438 & 137 & 0.19 & 0.92 & 4.98 \\
{\it Swift} & 0.0803 & 1357 & 0.72 & 3.84 & 5.32 \\
Thacher & 0.1200 & 186 & 0.09 & 0.42 & 4.54 \\
Las Cumbres & 0.1951 & 754 & 0.35 & 1.42 & 4.02 \\
Swope & 0.2459 & 204 & 0.59 & 1.87 & 3.19 \\
CSS & 4.9997 & 61 & 6.15 & 6.98 & 1.13 \\
PS1\tablenotemark{\footnotesize b} & 7.068 & 148 & 18.79 & 19.99 & 1.06 \\
GOTO-4 & 18.1300 & 399 & 30.48 & 32.01 & 1.05 \\
ATLAS & 28.8906 & 437 & 47.02 & 45.84 & 0.97 \\
ZTF & 46.7253 & 313 & 28.99 & 30.4 & 1.05 \\
\hline
{\bf All Tiles} & {\bf 9078.59}\tablenotemark{\footnotesize c} & {\bf 10984} & {\bf 48.13} & {\bf 48.28} & {\bf 1.00\tablenotemark{\footnotesize d}}
\enddata
\tablecomments{A synopsis of \teglon's effect on the community's combined EM search campaign for GW190425. \teglon~strongly enhances $\eta$ for instruments with FOVs $\leq1.0$~deg$^{2}$, see Section~\ref{sec:teglon_0425}.}
\vspace{-8pt}
\tablenotetext{a}{See Section~\ref{sec:teglon} for a detailed description.}
\vspace{-8pt}
\tablenotetext{b}{PS1 effective FOV is calculated from the average number of PS1 skycells in a fiducial PS1 FOV.}
\vspace{-8pt}
\tablenotetext{c}{Total unique area covered by all observations.}
\vspace{-8pt}
\tablenotetext{d}{Efficiency boost approaches 1 as both more area is observed and the FOV of the instrument increases; however, this relation is not monotonic because it depends on whether the original pointings remain in high-probability locales after \teglon~redistributes probability.}
\end{deluxetable*}

\subsection{Model Detection Efficiencies Calculated By \teglon}\label{sec:teglon_upper_limit}

A thorough presentation of \teglon's pixel math will be presented by Coulter et al. (2024, in prep.); however, a brief treatment here will serve to contextualize our results in Section~\ref{sec:model_comparison}. To compute the efficiency with which \teglon\ detects a model given a set of observations, or ``model detection efficiency,'' each instrument referenced in Section \ref{sec:data} is represented as a collection of polygons, and together with the celestial coordinates of every pointing in Table \ref{tab:observations}, uses the {\tt healpy} library to return every pixel contained within these observations from the final localization map for GW190425. Each of these pixels contain a marginal 2D probability for the GW originating from its sky position, $P_{\mathrm{2D},i}$, as well as GW-derived distance distribution parameters, $\bar{D_{i}}$ and $\sigma_{D_{i}}$.

For each pixel $i$, we then retrieve the set of covering $j$ observations (i.e., filter $f$ and limiting magnitude $m_{j,f}$), as well as the matching absolute magnitude for a model under consideration $M_{\mathrm{model}_{j,f}}$. We combine the line-of-sight extinction \citep[$A_{f}$; derived from][]{Schlafly11}, and reparameterize $m_{j,f}$ in terms of the distance $D_{\mathrm{model}_{j,f}}$ we would expect to detect a source in pixel $i$, as

\vspace*{-1.5em}
\begin{eqnarray}
\mu_{\mathrm{model}_{j,f}} =  m_{j,f} - M_{\mathrm{model}_{j,f}} - A_{f}\, , \\
D_{\mathrm{model}_{j,f}}~\mathrm{[Mpc]} = 10^{0.2 \times (\mu_{\mathrm{model}_{j,f}} - 25)}\, .
\end{eqnarray}\label{eqn:distance_weight}

\noindent To calculate the weight of finding the counterpart we integrate this distance distribution,

\begin{equation}
W_{\mathrm{model}_{i,j}} = \frac{1}{\sqrt{2\pi} \sigma_{D_i}} \int_0^{D_{\mathrm{model}_{j,f}}} e^{-\frac{1}{2} \left(\frac{\bar{D_{i}} - D}{\sigma_{D_{i}}}\right)} dD\, .
\end{equation}

To combine independent observations we take the complement of the joint probability that we do not see the source in {\it any} epoch --- that is, we weight the relative likelihood we would detect a specific model in image $j$ with $P_{\mathrm{2D}, i}$ and sum over all pixels to obtain a cumulative probability of detecting a specific model, 

\begin{equation}\label{epn:model_prob}
P_{\mathrm{model}} = \displaystyle\sum_{i}  P_{\mathrm{2D},i} \left[1 - \prod_{j} \left(1 - W_{\mathrm{model}_{i,j}}\right)\right]\, .
\end{equation}

\noindent This final model-detection efficiency is interpreted as the likelihood that we would have seen a source with the properties of the given model with our observations, for a wide range of models described below in Section~\ref{sec:model_comparison}. 

\section{Model Comparisons}\label{sec:model_comparison}

Based on the results of Section~\ref{sec:candidates_public}, we assume that there are no credible EM candidates for GW190425, and interpret the image depth for the data presented in Table \ref{tab:observations} and in the Treasure Map (described in Section \ref{sec:data}) as limits on a few classes of hypothetical EM counterparts to a BNS merger. To make this physically meaningful, we assume (and in the case of the data in Table \ref{tab:observations}, we {\it know}) that the data are homogeneous in that (1) each datum is the result of subtracting an in-band template image from the search image using the same instrument configuration, and (2) the reported depth of each image was computed by estimating the $\ge 3\sigma$ limiting magnitude from the difference image. We perform the joint model detection efficiency calculation, combining all reported epochs, depths, and filters, using the formalism described in Section \ref{sec:teglon_upper_limit}. The maximum probability to detect any model is limited by the total amount of probability that the full dataset covers; therefore, we report our detection probability in two ways: (1) as the probability calculated by Equation \ref{epn:model_prob}, and (2) as a percent of the total amount of redistributed probability reported in Table \ref{tab:search_synopsis}, 48.28\%, i.e., $X$\%~($\frac{X}{48.28}$\%). This relative detection efficiency characterizes the effectiveness of the observations themselves, assuming they could have covered the entire localization region.

\subsection{Kilonovae}\label{sec:model_comparison_kne}

The discovery of the EM counterpart to GW170817, the KN AT~2017gfo \citep{Coulter17, Abbott16:em}, demonstrated that BNS mergers are associated with short-lived thermal transients with luminous UVOIR emission, consistent with radioactive decay of freshly synthesized heavy elements \citep{Drout17}. A KN light curve's peak luminosity, color, and evolution timescale depend on the amount of mass the merger ejected $M_{\rm ej}$, the ejecta's expansion velocity $v_{\rm ej}$, and the ejecta's opacity $\kappa$ \citep{Arnett82}. The exact value of $\kappa$ is driven by the atomic structure of the specific chemical species, but in general, elements with atomic mass $A > 140$ have many millions of bound-bound line transitions so that their opacities are $> 10$ times that of Fe. This high opacity increases the photon diffusion timescale (and therefore the light curve evolution timescale), and shifts the emission from the UV/optical to the IR \citep{Kasen13}. Surprisingly, observations of AT~2017gfo showed that multiple ejecta components of different compositions (i.e., opacities) were required to accurately model its light curves \citep{Cowperthwaite17,Kilpatrick17,Villar17}. 

\begin{figure*}
    \centering
    \includegraphics[resolution=300,width=0.485\textwidth]{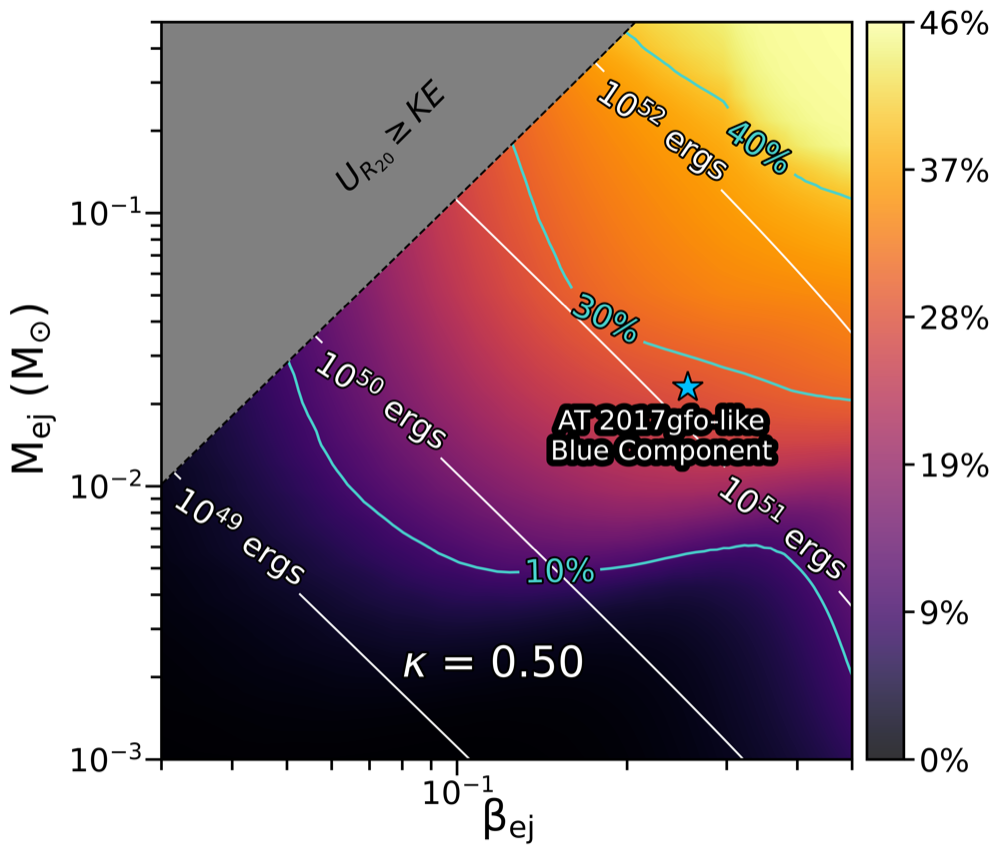}
    \includegraphics[resolution=300,width=0.495\textwidth]{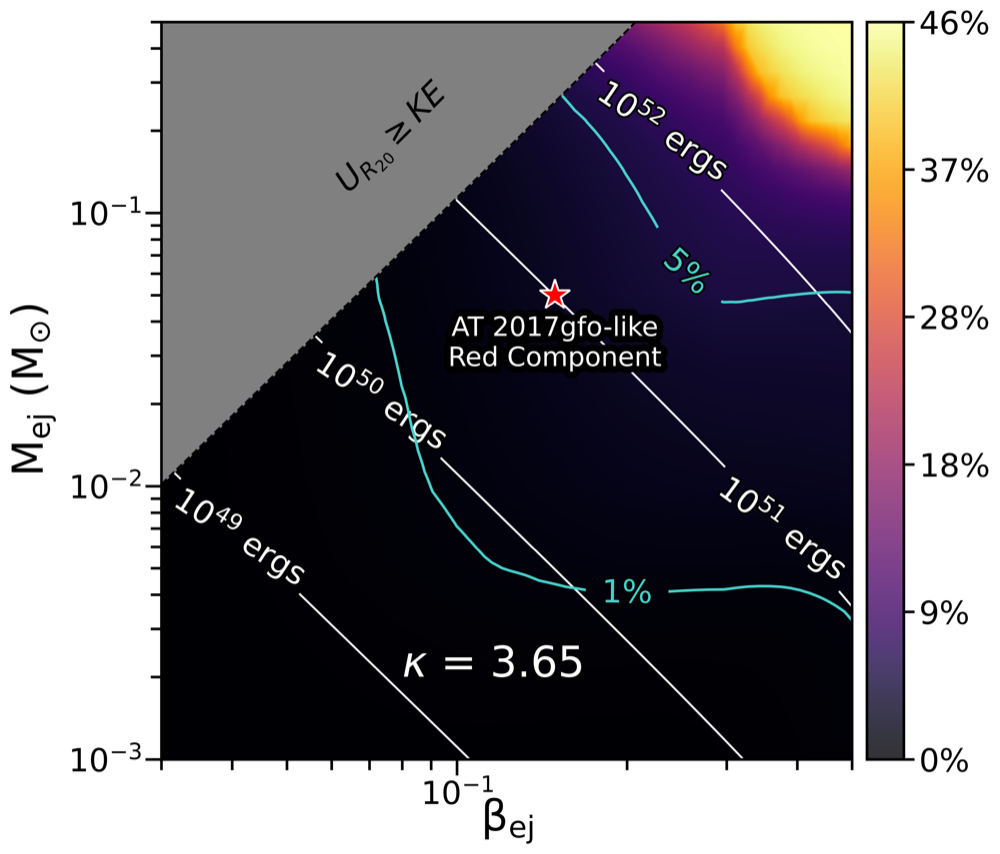}
    \caption{Detection probabilities for KN models from \citet{Villar17} as a function of ejecta mass ($M_{\rm ej}$ in M$_{\odot}$), ejecta velocity ($\beta_{\rm ej}$ in natural units), and opacity ($\kappa$). In the upper-left corner of each plot, we have grayed out the region where the binding energy of the ejecta mass exceeds its kinetic energy assuming a stiff NS equation of state and an NS radius of $20$~km. We show contours of equal probability in turquoise and contours of equal kinetic energy for the ejecta in white. Model values for AT~2017gfo-like components are taken from \citet{Villar17}, Table 2. {\it Left:} KN models with $\kappa=0.50$~cm$^2$~g$^{-1}$, overplotted with the blue component of AT~2017gfo ($\beta_{\rm ej}=0.256$, $M_{\rm ej} = 0.023$~M$_{\odot}$) which we can rule out at detection probability 28.4\%. {\it Right:} KN models with $\kappa=3.65$~cm$^2$~g$^{-1}$, overplotted with the red component of AT~2017gfo ($\beta_{\rm ej}=0.149$, $M_{\rm ej}=0.050$~M$_{\odot}$) which we cannot rule out at a detection probability of 2.9\%.}
    \label{fig:kilonova}
\end{figure*}

We consider two fiducial KNe models, a ``blue'' KN and a ``red'' KN, following the prescription of \citet{Villar17} to generate models with varying ejecta masses ($M_{\rm ej}$), velocities ($v_{\rm ej}$), and opacities \citep[see also][]{Villar17b}.  We do so in the Modular Open Source Fitter for Transients ({\tt MOSFiT}) framework \citep{mosfit}, with fixed parameters and bandpasses matched to our dataset.  In general, these models adopt a blackbody spectral profile with the photospheric temperature evolving to a floor, similar to the ``freeze out'' in spectral shape observed at late times with GW170817/AT~2017gfo \citep[$\sim 2500$~K; see][corresponding to the recombination temperature of species with open $f$ shells]{Drout17}.  We adopt $\kappa=0.5$~cm$^{2}$~g$^{-1}$ as the ``blue'' model and $\kappa=3.65~$cm$^{2}$~g$^{-1}$ as the ``red'' model, matching the parameters for a two-component model originally presented by \citet{Cowperthwaite17} and \citet{Villar17}.  Furthermore, in the context of an extremely lanthanide-rich model with high opacities, we consider $\kappa=10~$cm$^{2}$~g$^{-1}$, which is also the highest opacity model adopted by \citet{Villar17}.  Otherwise, we consider ejecta masses of 0.001--0.5~M$_{\odot}$ and velocities of 0.05$c$--0.50$c$ for all sets of models.  For each KN model set, we show our estimated detection probabilities in Figure~\ref{fig:kilonova}. We detect a blue, AT~2017gfo-like KN at 28.4\% (59.0\%) and are insensitive to a red, AT~2017gfo-like KN at 2.9\% (6.0\%). From these constraints, an immediate conclusion is that to be sensitive to more-massive events (e.g., another GW190425-like or NSBH event), EM search teams should search at redder wavelengths with deeper limits. See Section \ref{sec:discussion} for a discussion on coordinating multiband searches with \teglon.

\begin{figure}
    \centering
    \includegraphics[resolution=300,width=0.49\textwidth]{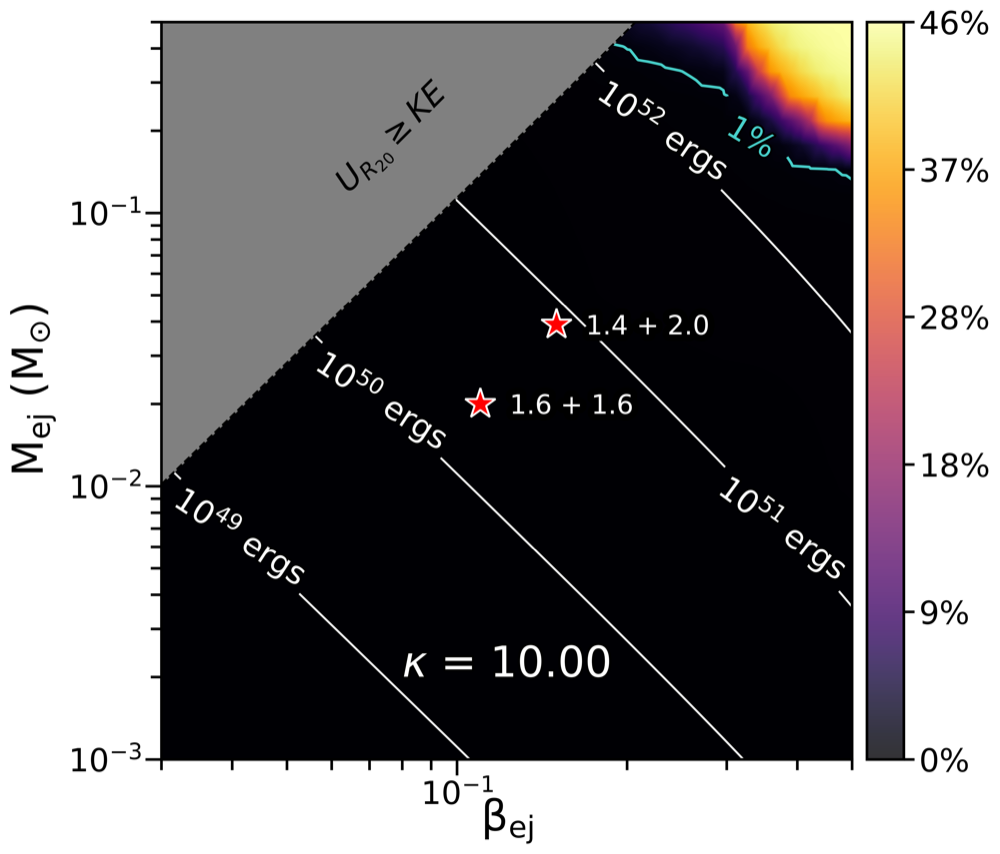}
    \caption{``Red'' KN models in the same style as Figure \ref{fig:kilonova}, with $\kappa=10.0$~cm$^2$~g$^{-1}$, chosen to reflect the high opacity of neutron-rich ejecta expected from a BNS merger that directly collapses to a BH. Overplotted are two realizations of ejecta mass and ejecta velocity of speculative KNe,
following the arguments of \citet{Foley20}, based on models from \citet{Rosswog13} that consider the case where the merging NSs have equal mass (1.6\,M$_{\odot}$ + 1.6\,M$_{\odot}$) or if one NS was more massive (1.4\,M$_{\odot}$ + 2.0\,M$_{\odot}$). We cannot rule out these red models given the depth of this combined dataset.}
    \label{fig:red_kilonova}
\end{figure}


\subsection{sGRB}\label{sec:model_comparison_sgrb}

We adopt an sGRB afterglow model {\tt JetFit} originally presented by \citet{Wu18} and \citet{Wu19}, and used to model the afterglow light curve of GRB\,170817A in the literature \citep[e.g.,][]{Hajela19,Hajela22,Kilpatrick22}.  For our fiducial model, we adopt the general parameters from the best fit to the multiwavelength GRB\,170817A light curve of \citet{Hajela22}.  These fixed parameters correspond to the electron energy fraction $\log \epsilon_{e}=-1$, the magnetic energy fraction $\log\epsilon_{B}=-5.17$, the spectral index of the electron distribution $p=2.15$, the asymptotic Lorentz factor $\eta_{0}=8.02$, and the boost Lorentz factor $\gamma_{B}=12$.  We then vary the explosion energy $E_0$, the ambient density $n$, and the viewing angle $\theta_{\rm obs}$ to generate in-band light curves from our fiducial model. We consider a range of ambient densities in units of particles per cm$^{3}$, ${\rm n} \in [10^{-6},10]$ cm$^{-3}$, and isotropic equivalent energy $E_{k,{\rm iso}}= 2 E_{0} / (1 - \cos(\frac{1}{2 \gamma_{B}}))$ in units of $10^{51}$~ergs (foe), $E_{k,{\rm iso}} \in [10^{-3},100]$ foe, consistent with observed sGRB afterglows in \citet{Fong15}. Finally, we considered two viewing angles ($\theta_{\rm obs}=0$ and $\theta_{\rm obs}=17^{\circ}$) for an ``on-axis'' and ``off-axis'' model (respectively), but report only on our relatively insensitive on-axis limits because our off-axis models are substantially fainter.  Our detection probability for a GRB~170918A-like model is $4.0 \times 10^{-1}$\% ($8.3 \times 10^{-1}$\%).


\begin{figure}
    \centering
    \includegraphics[resolution=300,width=0.49\textwidth]{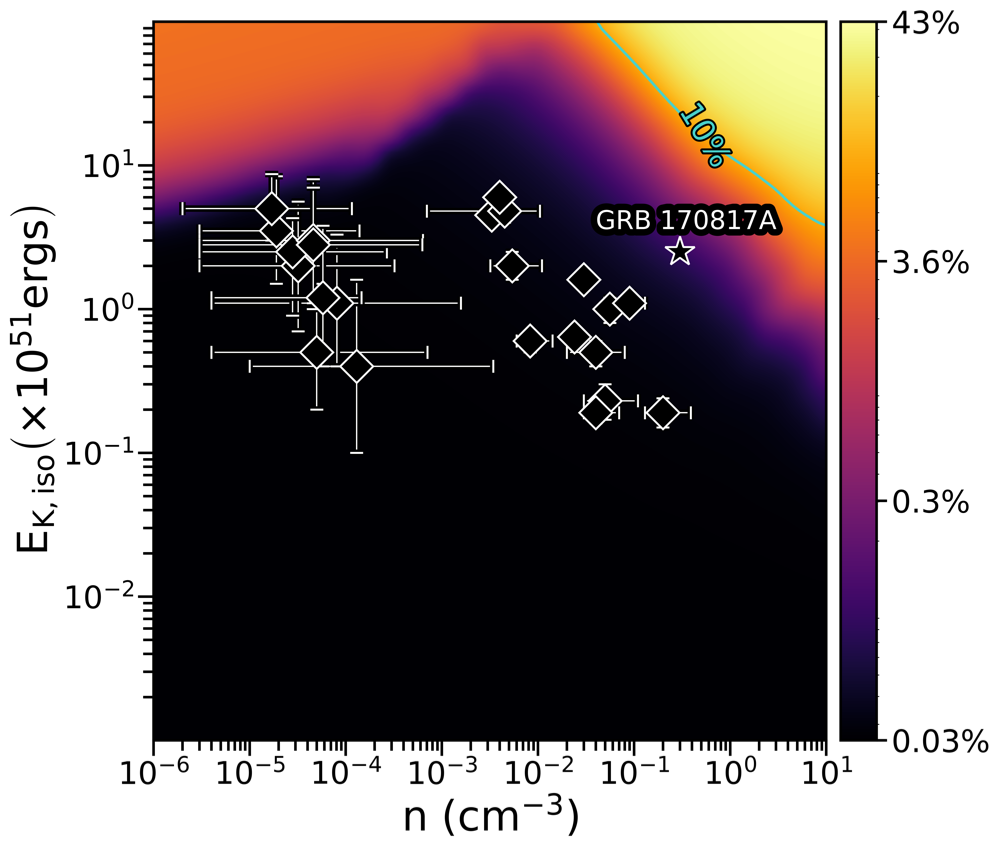}
    \caption{Limits on on-axis ($\theta_{\rm obs}=0^\circ$) sGRB models from \citet{Wu18}, as a function of isotropic kinetic energy ($E_{k,{\rm iso}}$ in foe) and circumburst density ($n$ in cm$^{-3}$). This dataset is relatively insensitive to these afterglow models, with a detection probability of a GRB~170817A-like model at $4.0 \times 10^{-1}$\% ($E_{k,{\rm iso}}=2.5$ foe, $n=0.3$~cm$^{-3}$, taken from \citet{Murguia-Berthier17}).}
    \label{fig:grb_onaxis}
\end{figure}

\subsection{Generic Models}\label{sec:model_comparison_generic}

KNe and sGRB afterglows have extremely short rise times, and it is likely that ground-based discoveries catch only their decline \citep[][]{Arcavi17, Drout17, Kilpatrick17}. Motivated by this, we include a generic class of empirical models parameterized by a peak absolute magnitude at the time of the merger, $M_{0}$, and a linear decline rate, $\Delta M$, in units of mag day$^{-1}$,

\begin{equation}
    M(t)=M_{0} -\Delta M (t-t_{0})\, ,
\end{equation}

\noindent where $t$ is in days. We make these models agnostic in their emission mechanism and construct their light curves with a flat spectral energy distribution (SED). Models are considered that span a peak magnitude range $M_{0} \in [-14,-20]$~mag and decline rates of $\Delta M \in [10^{-3},1.5]$~mag~day$^{-1}$ (in log space) to cover a parameter range that includes AT~2017gfo and several classes of well-known transients.

\begin{figure}
    \centering
    \includegraphics[resolution=300,width=0.49\textwidth]{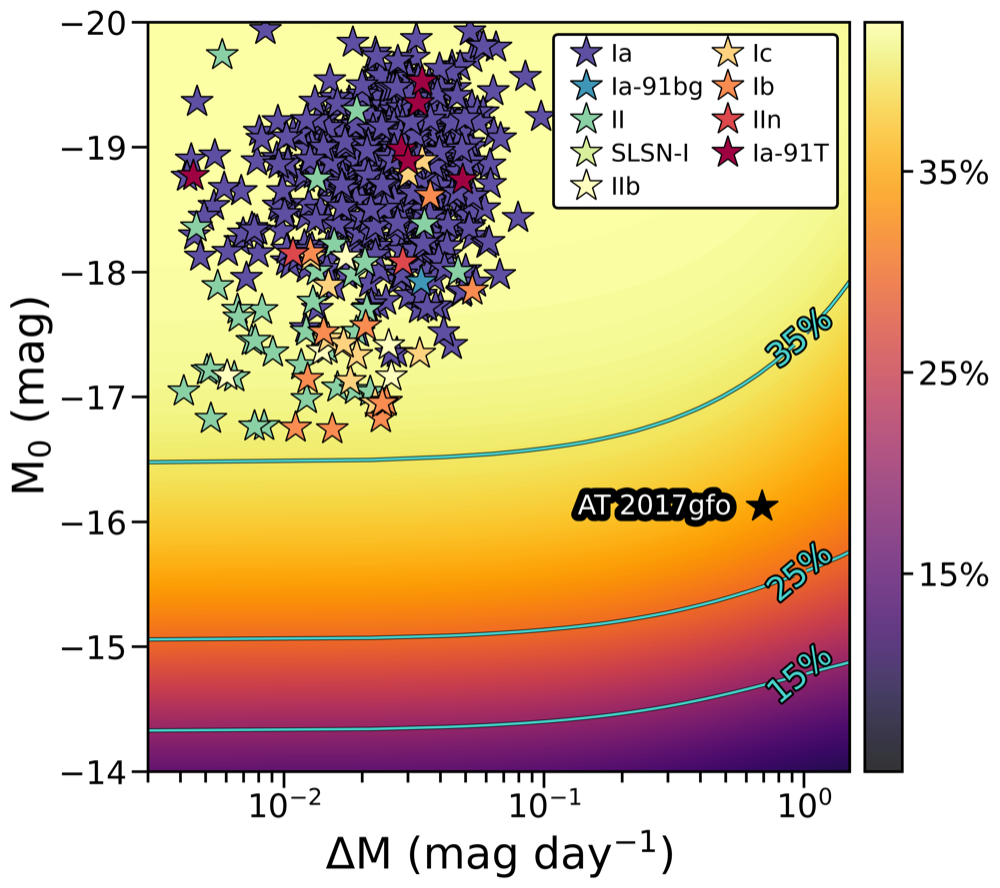}
    \caption{Limits on generic linear models parameterized by a peak absolute magnitude at the time of the merger ($M_{0}$ in mag) and a linear decline rate ($\Delta M$ in mag day$^{-1}$). Our models are agnostic in their emission mechanism and have a uniform SED. We overplot a range of well-known classes of transients taken from \citet{Siebert17}, as well as an average representation of AT~2017gfo \citep[$M\approx-16$ mag, $\Delta M\approx0.7$ mag day$^{-1}$; based on][]{Siebert17, Drout17, Kilpatrick17}. We rule out an AT~2017gfo-like counterpart with 30.0\% confidence.}
    \label{fig:linear}
\end{figure}

In Figure \ref{fig:linear}, we show our results, with parameters for AT~2017gfo representing an average of its decline across blue and red bands ($M\approx-16$ mag; $\Delta M\approx0.7$ mag day$^{-1}$) and a collection of transient types overplotted in juxtaposition \citep[referenced from][]{Siebert17}. To the limits of this dataset's coverage, we confidently detect these well-known extragalactic transient types, reinforcing the results reported in Section \ref{sec:candidates}. We rule out an AT~2017gfo-like model at $30.0$\% ($62.1$\%) confidence.

\section{Discussion}\label{sec:discussion}

As the second BNS merger identified by the LVC, GW190425 had significantly different source properties from the first BNS event GW170817.  In particular, the much larger total system mass of 3.4$\substack{+0.3\\-0.1}$~M$_{\odot}$ compared with 2.73$\substack{+0.04\\-0.01}$~M$_{\odot}$ \citep{Abbott17:detection} may lead to a significantly different EM counterpart \citep[see][]{Foley20}. Now that the LVK's O4 run is well underway, we are faced with a new paradigm for EM searches: more-distant events owing to an increase in detector sensitivity, larger than expected localization areas owing to Virgo's unexpected downtime during O4a, and an intrinsic diversity in BNS and NSBH systems leading to a range of EM counterparts. Therefore, it is likely that the search strategies that worked for GW170817 will have to be updated. To address these scenarios in O4 and beyond, we consider several updates to \teglon\ which will enhance existing capabilities and add new ones. Here we explore our plan to optimize a network of telescopes engaged in a counterpart search, add new catalogs to \teglon\ to enable new types of counterpart searches, and consider combining different types of localization information and source properties from coincident sources such as FRBs, GRBs, and neutrino detections into \teglon.

\subsection{EM Follow-up Coordination}

If we sum the product of every pixel with the multiplicity of observations that cover it and divide that number by two --- one epoch for a search image and one epoch for a template image --- these observations could have in principle uniquely covered $\sim 17,735$~deg$^{2}$, or roughly 1.8 times the final 90th percentile localization region. This ratio between the total area surveyed and 90th percentile localization region is even larger if we only consider the portions of the sky with no Sun constraint (i.e., the Eastern Spur of probability; see Figure \ref{fig:survey_footprint}). However, the EM community's follow-up strategy for GW sources is uncoordinated between observational teams, as evidenced by the search statistics shown in Table~\ref{tab:search_synopsis} --- only 5,638~deg$^{2}$ of localization area was uniquely searched. Furthermore, as seen in Figure \ref{fig:survey_footprint}, ZTF covered nearly the entire Eastern Spur of probability in $g+r$ over the course of the first 3 nights. Despite this, there were $\sim900$ other images taken of these same fields in the same filters within the same time period with 5 other instruments. In our analysis, these additional images offer little constraining power on the models that we consider. Despite the stroke of luck of GW170817 merging at a distance of only 40~Mpc, and its counterpart AT~2017gfo discovered just $\sim 11$~hr later, no KN was localized in O3 and the prospects for localizing one in O4 remain challenging. Increasing the coordination between follow-up facilities can drastically increase the odds of rapidly and precisely localizing the next KN by leveraging \teglon\ to design strategies that can optimize our sensitivity to a range of counterpart models.


To address these challenges, in an upcoming enhancement \teglon\ will publish its redistributed localization map as a dynamic, real-time service that can be subscribed to by a network of telescopes. For a given GW event, each telescope within the network will be incorporated into a global queue that will query \teglon\ for the next best observation (i.e., the next highest probability observation). When an observation is scheduled, \teglon\ will decrement the probability in the pixels that are covered, and the following query for the next best observation will be dynamically updated. This coordination function will operate on a per-filter basis, allowing different passbands to be optimized independently. Finally, the pixel probability decrementation will be dynamic: \teglon\ will alter the probability proportional to a model-specific light curve as a function of filter. For instance, while KNe quickly decline in blue bands, they rise more slowly in red bands. For joint searches in blue and red filters, \teglon\ will restore probability to covered pixels at different rates to force successive observations back to regions of high 2D probability depending on what filter the search instrument is using. In this way, \teglon\ will optimize a network of search telescopes in both their spatial coverage and model sensitivity.

\subsection{Specialized Catalog Additions: EM Counterparts to Binary Black Holes}

Immediately preceding O4, the LVK was expected to detect 260$\substack{+330\\-150}$ BBH mergers per year in O4\citep[see the International Gravitational-Wave Observatory Network Public Alerts User Guide\footnote{\url{https://emfollow.docs.ligo.org/userguide/capabilities.html\#summary-statistics}}, and ][]{Abbott:Local}, whereas the LVK discovered 75 BBH events in the 7.7~months of is O4a observing run\footnote{Not counting three BBH events in the engineering preceding O4a.}, a rate of 117~yr$^{-1}$ and consistent with the lower 1$\sigma$ bound of the expected rate.  While these BBH mergers are by themselves not expected to directly produce luminous transients, the nuclei of galaxies should host the densest populations of BH binaries, and for some of these binaries, they may be embedded in the disks of active galactic nuclei (AGNs). These environments can provide torques and tidal forces that can accelerate the pace of orbital decay \citep{Bartos17, Antoni19, Grobner20, Kaaz2023} and lead to mergers in baryon-rich environments. For such systems, BBH mergers may trigger AGN flares; such a transient is proposed to explain the optical flare discovered 34~days after the GW detection of the BBH merger GW190521 \citep[also known as S190521g;][although see \citet{Ashton20} for a rebuttal]{Graham20, Abbott20:GW190521}.

To facilitate the follow-up campaigns for BBHs, \teglon\ will be enhanced to include the AGN catalog from \citet{Secrest15}, which contains 1.4 million AGNs down to $g=26$~mag selected from the AllWISE catalog \citep{Wright2010}. This catalog is estimated to be complete for known AGNs to $\gtrsim 84$\%, and for all AGNs with $R<19$~mag.  Therefore, for AGNs with $z<0.1$, the catalog is expected to be close to $>90$\% complete. This catalog will provide an alternative galaxy catalog weighting scheme to accentuate AGN hosts within the LVK volume. 

\subsection{FRB~190425A and Combining Coincident Sources within \teglon}

While no optical counterparts were discovered in our follow-up campaign or imaging from other efforts that appear consistent with a KN or GRB afterglow from GW190425 \citep{Coughlin19, Hosseinzadeh19, Lundquist19, Antier20, Gompertz2020}, \citet{Moroianu23} reported the potential coincidence between FRB\,190425A and GW190425 based on the former's detection inside the 90th percentile credible region of the latter and discovery of the FRB 2.5~hr after the GW190425 merger.  Given their highly energetic radio bursts, millisecond timescales, and the discovery of an FRB from the Galactic magnetar SGR\,J1935+2154 \citep{CHIME-magnetar,Bochenek20, Zhang22}, FRBs are thought to arise from or in the immediate environments of magnetars \citep[see, e.g.,][for a discussion of various FRB emission models involving magnetars]{Margalit19,Metzger19b, Lyutikov+20}.  Invoking the formation of a magnetar in the post-merger collapse of a BNS system \citep[][]{Zhang13, Most18b}, FRBs may be credible radio counterparts to BNS mergers, and combining observables from GW events and FRBs within \teglon\ can aid in rapid localization and identification of likely host galaxies \citep[similar to the analysis of][]{Panther23}.

While arcsecond-scale localization of FRBs is possible with interferometers such as ASKAP \citep{Macquart10}, VLA \citep{Law18}, and MeerKAT \citep{Rajwade22}, the vast majority of FRBs are discovered with localizations of several deg$^{2}$ \citep[including FRB\,20190425A, e.g., by CHIME;][]{CHIME-DR1}.  At these angular scales, the \teglon\ algorithm is effective at selecting high-likelihood galaxies within the 2D localizations of both maps, for example by assuming that both the 2D localization provided by the LVK ($P_{\rm 2D,GW}$) and the CHIME beam ($P_{\rm 2D,FRB}$) represent independent estimates of the same source location and are combined into a single map ($P_{{\rm 2D}_{k}}=P_{\rm 2D,GW}\times P_{\rm 2D,FRB}$.  This assumption can be extended to any class of sources with localization on angular scales of degrees; indeed, the LVK produces combined skymaps incorporating localization information from third parties such as GRBs and neutrino alerts\footnote{See \url{https://emfollow.docs.ligo.org/userguide/content.html}.}.


\teglon\ can further benefit from FRB coincidences by incorporating distance constraints based on the dispersion measure (DM) obtained directly from the FRB signal.  This quantity correlates directly with the column of electrons along the line of sight to the FRB; combined with information on the density of electrons in the Milky Way, host-galaxy environments, and the intergalactic medium, this electron column density can constrain the distance to a FRB \citep{Deng14,Zhou14,Macquart20}.  In addition to multiple independent 2D localization constraints, the \teglon\ algorithm can accommodate multiple independent {\it volume} localizations by combining distance distributions within each map pixel, such as  replacing Equation~\ref{eqn:distance_weight} with a nonparametric distribution for each map pixel.

\section{Conclusions}

We present the most comprehensive analysis to date of UVOIR follow-up campaigns of the GW event GW190425 by the Gravity Collective and publicly reported data.

\begin{enumerate}
    \item We present new follow-up data from the Gravity Collective for GW190425, including optical and IR imaging from the KAIT, Nickel, Thacher, SMARTS 1.3\,m, and Swope telescopes covering a unique 54.76~deg$^{2}$ of the 90th percentile localization region and 3.99\% of the \teglon\ redistributed 2D probability (corresponding to 0.98\% of the LVK-assigned 2D probability) across {\em uBVgriIJHK} bands.  In addition, we present a new MOSFIRE IR spectrum of the high-probability candidate SN\,2019ebq, demonstrating that it is consistent with an SN\,Ib/c at $z=0.037$.

    \item We analyze all candidate counterparts discovered within the 90th-percentile localization region for GW190425, including their available spectra, possible identifications as minor planets or variable stars, host-galaxy associations and redshifts relative to the localization volume of GW190425, and photometry.  We are left with \Nremain\ candidates that we cannot rule out as being associated with GW190425, four of which we consider ``more likely'' candidates based on their time of discovery, host-galaxy associations, and implied luminosity.
    
    \item Assuming that none of these candidates is the counterpart to GW190425, we perform a joint analysis of our data combined with all publicly reported imaging using a new tool,  \teglon. This tool uses a 3D spatially varying galaxy catalog completeness weighting scheme, based on galaxy luminosity, to redistribute the original LVK 2D probability to account for local regions of high catalog completeness.  We have provided \teglon\footnote{\url{https://github.com/davecoulter/teglon_O4}}\citep[\coulterpub;][]{Coulter21} as an open-source tool available to the broader GW follow-up community.

    \item With \teglon, we homogeneously analyze this combined dataset, covering a unique 9,078.59 deg$^{2}$ and 48.28\% of the \teglon\ redistributed 2D probability (corresponding to 48.13\% of the LVK-assigned 2D probability) across UVOIR bands. We find that there was a 28.4\% and 2.9\% chance of detecting a KN similar to the blue and components (respectively) of AT~2017gfo in the combined dataset. Furthermore, we find that the data are generally insensitive to an on-axis sGRB, and rule out a generic transient with a similar peak luminosity and decline rate as AT~2017gfo to 30\% confidence.  Combining all new imaging data presented here as well as publicly available imaging in the literature, our \teglon\ analysis is the most comprehensive meta-analysis of GW190425 presented to date.
    
    \item Finally, we analyze the full search for optical counterparts to GW190425 in terms of the search strategy adopted across the astronomical community, unique optical counterparts such as those arising from NS mergers in the disks of AGNs, and the possible radio counterpart FRB\,190425A discovered 2.5~hr after the GW190425 merger.  We argue that \teglon\ can aid in each of these cases by optimally analyzing the search strategies for multiple telescopes with varying FOVs and depth, incorporating source catalogs apart from galaxies in its algorithm, and calculating the overlap between GW events and those from coincident events such as GRBs and FRBs into the localization maps that it generates.
\end{enumerate}

\software{{\tt astropy} \citep{astropy}, 
          {\tt DoPhot} \citep{Schechter93},
          {\tt dustmaps} \citep{Green18}, 
          {\tt healpy} \citep{healpy}, 
          {\tt hotpants} \citep{Becker15},
          {\tt ligo.skymap} \citep{GTD, GTDSupplement}, 
          {\tt PypeIt} \citep{pypeit:joss_arXiv, pypeit:zenodo},
          {\tt SExtractor} \citep{sextractor},
          \teglon~\citep{Coulter21},
          {\tt Treasure Map} \citep{Wyatt20}}

\facilities{KAIT, Keck:I (MOSFIRE), Nickel (Direct 2K), SMARTS 1.3m (ANDICAM), Swope (Direct 4K), Thacher (ACP).}

\begin{acknowledgments}

We appreciate the expert assistance of the staffs at the various observatories where data were obtained. We also acknowledge the important work that the Treasure Map is engaged in, by facilitating the public distribution of GW search data, which allowed this analysis to be conducted. 
D.A.C. thanks Leo Singer for many insightful conversations on topics as diverse as plotting to understanding GW posteriors through his comprehensive set of tools\footnote{\url{https://lscsoft.docs.ligo.org/ligo.skymap/}}, Michael Coughlin for many conversations about EM search optimizations, R.~H. for important conversations on the software engineering required to build \teglon, D.~O.~Jones for his technical expertise in quantifying our imaging depth, Antonella Palmese for her sharp eyes reviewing equations, Stephen Smartt and I.A. for graciously sharing their pointing data with our team, Aaron Tohuvavohu and Sam Wyatt for their help integrating Treasure Map queries into \teglon, and Ashley Villar for her insight into {\tt MOSFiT} modeling.


C.D.K. acknowledges partial support from a CIERA postdoctoral fellowship.
The work of P.M. was partly performed under the auspices of the U.S. Department of Energy by Lawrence Livermore National Laboratory under Contract DE-AC52-07NA27344. The document number is LLNL-JRNL-863367..
I.A. is grateful for support from the European Research Council (ERC) under the European Union’s Horizon 2020 research and innovation program (grant agreement 852097), from the Israel Science Foundation (grant 2752/19), from the United States -- Israel Binational Science Foundation (BSF; grant 2018166), and from the Pazy Foundation (grant 216312).
J.A.V. acknowledges the Doctorate in Astrophysics and Astroinformatics and Postgraduate School of the Universidad de Antofagasta for its support and allocated grants.
E.R.-R. is supported by the Heising-Simons Foundation and the NSF (AST-2307710, AST-2206243, AST-1911206, and AST-1852393).
J.S.B. is partially supported by the Gordon and Betty Moore Foundation and the NSF.
M.R.D. acknowledges support from the NSERC through grant RGPIN-2019-06186, the Canada Research Chairs Program, and the Dunlap Institute at the University of Toronto.
A.V.F.'s research group at UC Berkeley acknowledges                           
financial assistance from the Christopher R. Redlich Fund, as                 
 well as donations from Gary and Cynthia Bengier, Clark and                    
 Sharon Winslow, Alan Eustace (W.Z. is a Bengier-Winslow-Eustace
 Specialist in Astronomy), and many other individuals.

This work includes data obtained with the Swope telescope at Las Campanas Observatory, Chile, as part of the Swope Time Domain Key Project (PI Piro; CoIs Burns, Coulter, Cowperthwaite, Dimitriadis, Drout, Foley, French, Holoien, Hsiao, Kilpatrick, Madore, Phillips, and Rojas-Bravo).
This work makes use of data from the Las Cumbres Observatory global telescope network. The LCO group is supported by NSF grants AST-1911225 and AST-1911151, and BSF grant 2018166.
This research has used data from the SMARTS 1.3\,m telescope, which is operated as part of the SMARTS Consortium.
Some of the data presented herein were obtained at the W. M. Keck Observatory, which is operated as a scientific partnership among the California Institute of Technology, the University of California, and NASA. The Observatory was made possible by the generous financial support of the W. M. Keck Foundation.
The authors wish to recognize and acknowledge the very significant cultural role and reverence that the summit of Maunakea has always had within the indigenous Hawaiian community.  We are most fortunate to have the opportunity to conduct observations from this mountain.
Research at Lick Observatory is partially supported by a generous gift        
from Google.  KAIT and its ongoing operation were made possible by            
donations from Sun Microsystems, Inc., the Hewlett-Packard Company,           
AutoScope Corporation, Lick Observatory, the NSF, the University of           
California, the Sylvia and Jim Katzman Foundation, and the TABASGO            
Foundation. 
We also appreciate data \lcogtpub\ shared by I.A. on behalf of Las Cumbres Observatory.

\end{acknowledgments}

\bibliography{gw190425}

\appendix


\startlongtable
\begin{deluxetable}
        {lcccccc}
        \tabletypesize{\scriptsize}
        \tablecaption{1M2H UVOIR Imaging of the GW190425 Localization Region\label{tab:observations}}
        \tablewidth{0pt}
        \tablehead{
        \colhead{Source\tablenotemark{\footnotesize a}} & 
        \colhead{$\alpha$} & 
        \colhead{$\delta$} & 
        \colhead{Exposure Time} &
        \colhead{Date\tablenotemark{\footnotesize b}} &
        \colhead{Filter} &
        \colhead{Magnitude Limit\tablenotemark{\footnotesize c}} \\
        &
        (J2000) &
        (J2000) &
        (s) &
        (MJD) &
        &
        (3$\sigma$)
        }
        \startdata
ANDICAM CCD & 17:31:09.72 & -08:27:14.40 & 900 & 58599.2453 & $I$ & 21.60 \\
ANDICAM CCD & 16:52:07.06 & -17:03:27.72 & 240 & 58599.2798 & $I$ & 19.91 \\
ANDICAM CCD & 15:40:36.10 & +28:30:30.60 & 240 & 58599.2895 & $I$ & 20.24 \\
ANDICAM CCD & 15:41:53.64 & +28:07:50.52 & 240 & 58599.2938 & $I$ & 19.60 \\
ANDICAM CCD & 15:47:53.42 & +25:43:31.80 & 240 & 58599.2984 & $I$ & 21.06 \\
ANDICAM CCD & 15:56:03.36 & +24:26:40.56 & 240 & 58599.3027 & $I$ & 20.81 \\
ANDICAM CCD & 15:57:13.85 & +25:48:54.36 & 240 & 58599.3074 & $I$ & 20.87 \\
ANDICAM CCD & 16:02:59.59 & +27:00:23.40 & 240 & 58599.3125 & $I$ & 21.15 \\
\enddata
        
    \tablecomments{We only include data that 1M2H acquired and reduced. For LCO data referred to in \autoref{sec:data}, these data will be published in \lcogtpub. For all other data, see the curated pointings on the Treasure Map \citep{Wyatt20}.}
    \tablenotetext{a}{Imaging as described in \autoref{sec:data}.}
    \tablenotetext{b}{MJD is taken from the center of the exposure time.}
    \tablenotetext{c}{In-band 3$\sigma$ limit for the reported image as described in \autoref{sec:data} and \autoref{sec:model_comparison}. All magnitudes are on the AB system \citep{Oke83}.}
    \end{deluxetable}

\startlongtable
\begin{deluxetable*}
{lccccccl}
\tabletypesize{\scriptsize}
\tablecaption{Candidate Electromagnetic Counterparts to GW190425\label{tab:candidates}}
\tablewidth{0pt}
\tablehead{
\colhead{Name}& 
\colhead{$\alpha$}& 
\colhead{$\delta$}& 
\colhead{Cumulative}&
\colhead{Discovery Date}&
\colhead{Redshift\tablenotemark{\footnotesize  b}}&
\colhead{Absolute Magnitude\tablenotemark{\footnotesize  c}}& 
\colhead{Note\tablenotemark{\footnotesize d}}\\[-0.25cm]
&
\colhead{(J2000)} &
\colhead{(J2000)} &
\colhead{LVC Prob.\tablenotemark{\footnotesize  a}}&
\colhead{(MJD)} &
&
\colhead{(AB mag)} &
}
\startdata
2019ocg  & 16:52:45.011 & -19:05:38.88 & 0.0923 & 58598.37600 & 0.08873$\pm$0.01265 (PS1)                      & $r$=$-$18.97$\pm$0.41 (0.1 d)    & (Z)                                  \\
2019hae  & 16:55:14.932 & -17:52:54.37 & 0.0768 & 58598.37600 & 0.07071$\pm$0.00281 (PS1)                      & $r$=$-$20.53$\pm$0.40 (0.1 d)    & (Z)                                  \\
2019ocf  & 19:14:46.398 & -03:00:27.02 & 0.8576 & 58598.38000 & --                                             & --                                & (PRE; $r$=-20.31$\pm$0.43, -166.11 d)\\
2019dzg  & 13:18:15.181 & +29:40:12.98 & 0.8487 & 58598.38100 & 0.44779$\pm$0.09683 (PS1)                      & $r$=$-$22.76$\pm$0.44 (0.0 d)    & (Z)                                  \\
2019oce  & 16:16:19.969 & +21:44:27.38 & 0.0044 & 58598.39300 & 0.11777$\pm$0.01048 (PS1)                      & $r$=$-$20.97$\pm$0.40 (0.9 d)    & (Z)                                  \\
2019eao  & 13:01:18.635 & +52:09:02.15 & 0.3742 & 58598.39699 & --                                             & --                                &                                      \\
2019eib  & 16:52:39.451 & +10:36:08.26 & 0.1341 & 58598.39800 & 0.14500$\pm$0.00145 (s)                        & $r$=$-$19.75$\pm$0.45 (-2.0 d)   & (SN) Ia                              \\
2019ebv  & 15:02:41.207 & +29:12:01.90 & 0.6153 & 58598.40300 & 0.20530$\pm$0.06050 (PS1)                      & $r$=$-$19.16$\pm$0.44 (0.1 d)    & (Z)                                  \\
\enddata

\tablecomments{We show every candidate considered in the analysis described in Section~\ref{sec:candidates}.  Absolute magnitudes are only shown for photometry that we use to rule out candidate counterparts as described in our analysis.}
\tablenotetext{a}{Cumulative probability within the {\tt GW190425\_skymap} provided by \citet{Abbott20:gw190425}.}
\tablenotetext{b}{Redshifts are identified as spectroscopic (s) or photometric (by source).}
\tablenotetext{c}{We indicate the absolute magnitude at our preferred redshift or the corresponding pixel in the final {\tt GW190425\_skymap} of observations used to rule out this candidate as being associated with GW190425.  Where no magnitude was provided, we do not find any photometry that is inconsistent with one of our photometric models.}
\tablenotetext{d}{We rule out each source based on classification as likely minor planets (MP), supernovae (SN), pre-merger variability (VAR or PRE), a redshift inconsistent with the GW190425 volume (Z), or photometric evolution (PHOT) as described in \autoref{sec:candidates}.}
\end{deluxetable*}

\section{Detailed Candidate Analysis}\label{app:candidates}

In Table~\ref{tab:candidates} we classify each candidate and indicate what criteria were used to rule out its association with GW190425, following methods similar to those of \citet{Kilpatrick21}.  Of the \Nremain\ remaining candidates that we cannot rule out, we describe what is known about each source and whether it could be a viable electromagnetic counterpart to GW190425.

Several other analyses have presented a discussion of some subset of the candidate optical counterparts to GW190425 that we consider here \citep[e.g.,][]{Coughlin19,Hosseinzadeh19,Lundquist19,Rastinejad22,Paek23}.  We note that of these publications, only \citet{Rastinejad22} report candidates that remain ``viable'' as counterparts to GW190425 after the cuts performed in their analysis, specifically AT\,2019efr and AT\,2019eig.


While we cannot rule out any of these candidates, many are unlikely to be counterparts based on some reasonable assumptions.  Any candidate that has no associated host galaxy corresponding to $M < -13$~mag for the distance to GW190425 and the typical depth for optical surveys, and was discovered $>5$ days after merger, is considered unlikely to be the counterpart.  Instead, these are likely high-redshift interlopers.  This results in four ``more likely'' candidates (AT\,2019ean, 2019ego, 2019egj, and 2019aasq) that we discuss below. Finally, we conclude our analysis by considering the two high-probability counterparts reported by \citet{Rastinejad22}.

\subsection{Candidate Optical Counterparts to GW190425}\label{sec:candidate_counterparts}

{\it 2019ean}: AT\,2019ean was discovered by ZTF 0.11~days after the GW190425 merger with an initial brightness of $r=19.87~mag$ and 20\arcsec\ (12~kpc) from its likely host galaxy IC~4611 at $z = 0.029841$ \citep{2019TNSTR.634....1F}.  As shown in Table~\ref{tab:candidates}, it is located at the 38.8th cumulative percentile most likely part of the final GW190425 map.  No additional forced-photometry detections of AT\,2019ean were recovered by ATLAS or ZTF despite significant coverage by both surveys within $\pm 7$~days of discovery.  It is therefore likely that AT\,2019ean peaked at $M_{r} \approx -15.8$~mag at the distance of its host galaxy within the first half day of discovery, close to models of AT\,2017gfo \citep[e.g., in][]{Cowperthwaite17,Drout17,Kasliwal17,Kilpatrick17,Smartt17,Villar17}.  We therefore consider AT\,2019ean to be a candidate kilonova counterpart to GW190425.

{\it 2019ego}: AT\,2019ego was discovered by {\it Gaia} 1.80~days after the GW190425 merger with an initial brightness of $G=18.97\pm0.20$~mag and 5.1\arcsec\ (3~kpc) from its likely host galaxy WISEA J004046.31-512807.8 at $z=0.032139$ \citep{2019TNSTR.662....1D}.  It is located at the 24.4th cumulative percentile most likely part of the final GW190425 map. Owing to its southern sky location at $\delta=-51.46831$ (J2000), there were limited follow-up observations at this position, hence precluding a detailed photometric classification of this event.  Regardless, AT\,2019ego was likely close to its peak magnitude of $M_{G}=-16.8$~mag around the time of discovery, and so it remains a candidate kilonova counterpart to GW190425 given its localization and known photometry.

{\it 2019egj}: AT\,2019egj was discovered by the MASTER survey 3.59~days after the GW190425 merger with an initial brightness of $19$~mag in a Clear filter and 12.4\arcsec\ (3~kpc) from the center of its likely edge-on spiral host galaxy SDSS J142814.30+304257.4 at $z=0.012784$ \citep{2019TNSTR.662....1D}.  It is located at the 36.5th cumulative percentile most likely part of the final GW190425 map.  Despite significant coverage by ATLAS and ZTF within $\pm 7$~days of the event, no other detections were obtained to similar depths as the discovery magnitude, and so AT\,2019egj was likely close to its peak brightness of $M_{\rm Clear}=-15$~mag at the time of discovery.  AT\,2019egj therefore remains a candidate kilonova counterpart to GW190425 given its localization and known photometry.

{\it 2019aasq}: AT\,2019aasq was discovered by the 1M2H collaboration using the Swope 1\,m telescope +4.93~days after the GW190425 merger with an initial brightness of $i=21.30\pm0.19$~mag and 7.5\arcsec\ from the center of its likely host galaxy WISEA J154032.14+282013.7 at $z=0.031090$ \citep{2023TNSTR1610....1C}, placing it close to $M_{i}=-14.4$~mag at the time of discovery.  It is located at the 96.9th cumulative percentile most likely part of the final GW190425 map.  Although ATLAS and ZTF had significant coverage of this region within $\pm 7$~days of the discovery, there are no other detections of this source, indicating that it was likely faint across most optical bands.  We therefore consider AT\,2019aasq as a candidate kilonova counterpart to GW190425.

We emphasize that while we consider these candidates to be ``more likely'' kilonovae, the implied ejecta masses based on their discovery magnitudes are large even compared with numerical relativity simulations using the component masses of GW190425.  For example, in the low-spin prior scenario ($\chi<0.05$), the expected component masses for GW190425 are $\sim 1.7$~M$_{\odot}$ and $\sim 1.6$~M$_{\odot}$ \citep{Abbott20}, which is expected to promptly collapse to a black hole and yield $<0.01$~M$_{\odot}$ of ejecta based on simulations from \citet{Radice18}.  The transients we investigate would all require ejecta masses $>0.07$~M$_{\odot}$ assuming a fixed ejecta velocity of $0.1c$ and $\kappa=3.0$~cm$^{2}$~g$^{-1}$, and using the models explored in Section~\ref{sec:model_comparison_kne}.  Thus, while we consider these transients more likely to be counterparts to GW190425 than all other candidates in our analysis, we also consider them unlikely to be kilonovae.

The following two high-probability counterparts were reported by \citet{Rastinejad22} as being viable counterparts to GW190425.

{\it 2019efr}: AT\,2019efr has a high-probability association with a Legacy DR10 galaxy at $\alpha=246.73420^{\circ}$, $\delta=10.93818^{\circ}$ (J2000) whose photometric redshift is 0.38057$\substack{+0.32302\\-0.21038}$, placing it outside the likely GW190425 volume.

{\it 2019eig}: We consider AT\,2019eig to be too bright at +4.3~days from merger to be a viable optical counterpart to GW190425. It has a {\it Gaia} $G$-band absolute magnitude of $-18.24\pm0.45$ based on its discovery magnitude, assuming that it is located at the distance reported by its corresponding pixel in the GW190425 map \citep{2019eig-disc}.


\end{document}